%%%%%%%%%%%%%%%%%%%%%%%%%%%%%%%%%%%%%%%%%%%%%%%%%%%%%%%%%%%%%%%%%%%%%%%%%%
%                                                                        %
%                                                                        % 
%                         by E. Guitter                                  %
%                TEX file, using lanlmac.tex macros                      %
%                                                                        %
%                                                                        %
%%%%%%%%%%%%%%%%%%%%%%%%%%%%%%%%%%%%%%%%%%%%%%%%%%%%%%%%%%%%%%%%%%%%%%%%%%
\input lanlmac
\def\href#1#2{{#2}}
\def\hhref#1{{#1}}
\input epsf.tex

\overfullrule=0mm

\newcount\figno
\figno=0
\def\fig#1#2#3{
\par\begingroup\parindent=0pt\leftskip=1cm\rightskip=1cm\parindent=0pt
\baselineskip=11pt
\global\advance\figno by 1
\midinsert
\epsfxsize=#3
\centerline{\epsfbox{#2}}
\vskip 12pt
{\bf Fig.\ \the\figno:} #1\par
\endinsert\endgroup\par
}
\def\figlabel#1{\xdef#1{\the\figno}}
\def\encadremath#1{\vbox{\hrule\hbox{\vrule\kern8pt\vbox{\kern8pt
\hbox{$\displaystyle #1$}\kern8pt}
\kern8pt\vrule}\hrule}}

%Macros 
%%%%%%%%%%%%%%%%%%%%%%%%%%%%%%%%%%%%%%%%%%%%%%%%%%%%%%%%%%%%%%%%%

\def\IR{\relax{\rm I\kern-.18em R}}
\font\cmss=cmss10 \font\cmsss=cmss10 at 7pt

\font\cmss=cmss10 \font\cmsss=cmss10 at 7pt
\def\IZ{\relax\ifmmode\mathchoice
{\hbox{\cmss Z\kern-.4em Z}}{\hbox{\cmss Z\kern-.4em Z}}
{\lower.9pt\hbox{\cmsss Z\kern-.4em Z}}
{\lower1.2pt\hbox{\cmsss Z\kern-.4em Z}}\else{\cmss Z\kern-.4em Z}\fi}
\def\IN{\relax{\rm I\kern-.18em N}}
\def\b{\circ}
\def\n{\bullet}

\def\gbbbb{\Gamma_4^{\hbox{$\scriptstyle \b \b$}\kern -8.2pt
\raise -4pt \hbox{$\scriptstyle \b \b$}}}
\def\gnnnn{\Gamma_4^{\hbox{$\scriptstyle \n \n$}\kern -8.2pt  
\raise -4pt \hbox{$\scriptstyle \n \n$}}}
\def\gnnnnnn{\Gamma_6^{\hbox{$\scriptstyle \n \n \n$}\kern -12.3pt
\raise -4pt \hbox{$\scriptstyle \n \n \n$}}}
\def\gbbbbbb{\Gamma_6^{\hbox{$\scriptstyle \b \b \b$}\kern -12.3pt
\raise -4pt \hbox{$\scriptstyle \b \b \b$}}}
\def\gbbbbc{\Gamma_{4\, c}^{\hbox{$\scriptstyle \b \b$}\kern -8.2pt
\raise -4pt \hbox{$\scriptstyle \b \b$}}}
\def\gnnnnc{\Gamma_{4\, c}^{\hbox{$\scriptstyle \n \n$}\kern -8.2pt
\raise -4pt \hbox{$\scriptstyle \n \n$}}}
\def\Rbud#1{{\cal R}_{#1}^{-\kern-1.5pt\blacktriangleright}}
\def\Rleaf#1{{\cal R}_{#1}^{-\kern-1.5pt\vartriangleright}}
\def\Rbudb#1{{\cal R}_{#1}^{\circ\kern-1.5pt-\kern-1.5pt\blacktriangleright}}
\def\Rleafb#1{{\cal R}_{#1}^{\circ\kern-1.5pt-\kern-1.5pt\vartriangleright}}
\def\Rbudn#1{{\cal R}_{#1}^{\bullet\kern-1.5pt-\kern-1.5pt\blacktriangleright}}
\def\Rleafn#1{{\cal R}_{#1}^{\bullet\kern-1.5pt-\kern-1.5pt\vartriangleright}}
\def\Wleaf#1{{\cal W}_{#1}^{-\kern-1.5pt\vartriangleright}}
\def\Cleaf{{\cal C}^{-\kern-1.5pt\vartriangleright}}
\def\Cbud{{\cal C}^{-\kern-1.5pt\blacktriangleright}}
\def\Crleaf{{\cal C}^{-\kern-1.5pt\circledR}}

%%%%%%%%%%%%%%%%%%%%%%%%%%%%%%%%%%%%%%%%%%%%%%%%%%%%%%%%%%%%%%%%%
%%%%%%%%%%%%%%%%%%%%%%%%%%%%%%%%%%%%%%%%%%%%%%%%%%%%%%%%%%%%%%%%%%%%%

\magnification=\magstep1
\baselineskip=12pt
\hsize=6.3truein
\vsize=8.7truein
 at 8truept
 at 8truept
 at 10truept
%\footline={\footsc the electronic journal of combinatorics
%   {\footbf 11} (2004), \#R00\hfil\footrm\folio}

%%%%%%%%%%%%%%%%%%%%%%%%%%%%%%%%%%%%%%%%%%%%%%%%%%%%%%%%%%%%%%%%%%%%%%%%
\font\bigrm=cmr12 at 14pt
\centerline{\bigrm Distance statistics in large toroidal maps}

\bigskip\bigskip

\centerline{E. Guitter}
  \smallskip
  \centerline{Institut de Physique Th\'eorique}
  \centerline{CEA, IPhT, F-91191 Gif-sur-Yvette, France}
  \centerline{CNRS, URA 2306}
\centerline{\tt emmanuel.guitter@cea.fr}

  \bigskip

     \bigskip\bigskip

     \centerline{\bf Abstract}
     \smallskip
     {\narrower\noindent
We compute a number of distance-dependent universal scaling functions 
characterizing the distance statistics of large maps of genus one. In
particular, we obtain explicitly the probability distribution for the 
length of the shortest non-contractible loop passing via a random point
in the map, and that for the distance between two random points. Our 
results are derived in the context of bipartite toroidal quadrangulations, 
using their coding by well-labeled $1$-trees, which are maps of genus one 
with a single face and appropriate integer vertex labels. Within this
framework, the distributions above are simply obtained as scaling limits 
of appropriate generating functions for well-labeled $1$-trees, all 
expressible in terms of a small number of basic scaling functions for 
well-labeled plane trees. 
\par}

     \bigskip

%references
\nref\TUT{W. Tutte,
{\it A Census of planar triangulations} Canad. J. of Math. {\bf 14} (1962) 21-38;
{\it A Census of Hamiltonian polygons} Canad. J. of Math. {\bf 14} (1962) 402-417;
{\it A Census of slicings}, Canad. J. of Math. {\bf 14} (1962) 708-722;
{\it A Census of Planar Maps}, Canad. J. of Math. {\bf 15} (1963) 249-271.
}
\nref\BIPZ{E. Br\'ezin, C. Itzykson, G. Parisi and J.-B. Zuber, {\it Planar
Diagrams}, Comm. Math. Phys. {\bf 59} (1978) 35-51.}
\nref\PDF{P. Di Francesco, {\it 2D Quantum Gravity, Matrix Models and Graph 
Combinatorics}, Lecture notes given at the summer school ``Applications of 
random matrices to physics'', Les Houches, June 2004, arXiv:math-ph/0406013.}
\nref\CORV{R. Cori and B. Vauquelin, {\it Planar maps are well labeled trees},
Canad. J. Math. {\bf 33(5)} (1981) 1023-1042.}
\nref\SThesis{G. Schaeffer, {\it Conjugaison d'arbres
et cartes combinatoires al\'eatoires}, PhD Thesis, Universit\'e 
Bordeaux I (1998).} 
\nref\MOB{J. Bouttier, P. Di Francesco and E. Guitter. {\it 
Planar maps as labeled mobiles},
Elec. Jour. of Combinatorics {\bf 11} (2004) R69, arXiv:math.CO/0405099.}
\nref\QGRA{V. Kazakov, {\it Bilocal regularization of models of random
surfaces}, Phys. Lett. {\bf B150} (1985) 282-284; F. David, {\it Planar
diagrams, two-dimensional lattice gravity and surface models},
Nucl. Phys. {\bf B257} (1985) 45-58; J. Ambj\o rn, B. Durhuus and J. Fr\"ohlich,
{\it Diseases of triangulated random surface models and possible cures},
Nucl. Phys. {\bf B257} (1985) 433-449; V. Kazakov, I. Kostov and A. Migdal
{\it Critical properties of randomly triangulated planar random surfaces},
Phys. Lett. {\bf B157} (1985) 295-300.}
\nref\DGZ{for a review, see: P. Di Francesco, P. Ginsparg and 
J. Zinn--Justin, {\it 2D Gravity and Random Matrices},
Physics Reports {\bf 254} (1995) 1-131.}
\nref\Duplantier{see for instance B. Duplantier, 
{\it Higher Conformal Multifractality},
J. Stat. Phys. {\bf 110} 691-738 (2003), 
arXiv:cond-mat/0207743.}
\nref\BauerBernard{see also M. Bauer and D. Bernard, 
{\it Conformal Field Theories of 
Stochastic Loewner Evolutions}, Commun. Math. Phys. {\bf 239} (2003) 493-521,
arXiv:hep-th/0210015. }
\nref\MARMO{J. F. Marckert and A. Mokkadem, {\it Limit of normalized
quadrangulations: the Brownian map}, Annals of Probability {\bf 34(6)}
(2006) 2144-2202, arXiv:math.PR/0403398.}
\nref\LEGALL{J. F. Le Gall, {\it The topological structure of scaling limits 
of large planar maps}, invent. math. {\bf 169} (2007) 621-670,
arXiv:math.PR/0607567.}
\nref\LGP{J. F. Le Gall and F. Paulin,
{\it Scaling limits of bipartite planar maps are homeomorphic to the 2-sphere},
Geomet. Funct. Anal. {\bf 18}, 893-918 (2008), arXiv:math.PR/0612315.}
\nref\MierS{G. Miermont, {\it On the sphericity of scaling limits of 
random planar quadrangulations}, Elect. Comm. Probab. {\bf 13} (2008) 248-257, 
arXiv:0712.3687 [math.PR].}
\nref\LEGALLGEOD{J.-F. Le Gall, {\it Geodesics in large planar maps and 
in the Brownian map}, Acta Math., to appear, arXiv:0804.3012 [math.PR].}
\nref\AW{J. Ambj\o rn and Y. Watabiki, {\it Scaling in quantum gravity},
Nucl.Phys. {\bf B445} (1995) 129-144, arXiv:hep-th/9501049.}
\nref\AJW{J. Ambj\o rn, J. Jurkiewicz and Y. Watabiki,
{\it On the fractal structure of two-dimensional quantum gravity},
Nucl.Phys. {\bf B454} (1995) 313-342, arXiv:hep-lat/9507014.}
\nref\ADJ{see also J. Ambj\o rn, B. Durhuus and T. Jonsson, 
{\it Quantum Geometry:
A statistical field theory approach}, Cambridge University Press, 1997.}
\nref\GEOD{J. Bouttier, P. Di Francesco and E. Guitter, {\it Geodesic
distance in planar graphs}, Nucl. Phys. {\bf B663}[FS] (2003) 535-567, 
arXiv:cond-mat/0303272.}
\nref\THREEPOINT{J. Bouttier and E. Guitter, {\it The three-point function 
of planar quadrangulations}, J. Stat. Mech. (2008) P07020, 
arXiv:0805.2355 [math-ph].}
\nref\LOOP{J. Bouttier and E. Guitter, {\it Confluence of geodesic paths and 
separating loops in large planar quadrangulations}, 
J. Stat. Mech. (2009) P03001, arXiv:0811.0509 [math-ph].}
\nref\PSEUDO{J. Bouttier and E. Guitter 
{\it Distance statistics in quadrangulations with a boundary, or with a 
self-avoiding loop}, J. Phys. A: Math. Theor. {\bf 42} (2009) 465208, 
arXiv:0906.4892 [math-ph].}
\nref\MS{M. Marcus and G. Schaeffer, {\it Une bijection simple pour les
cartes orientables} (2001), available at 
\hhref{http://www.lix.polytechnique.fr/Labo/Gilles.Schaeffer/Biblio/}.}
\nref\CMS{see also G. Chapuy, M. Marcus and G. Schaeffer, 
{\it A bijection for rooted maps on orientable surfaces}, 
SIAM J. Discrete Math. {\bf 23}(3) (2009) 1587-1611, 
arXiv:0712.3649 [math.CO].}
\nref\Bett{J. Bettinelli, {\it Scaling Limits for Random Quadrangulations 
of Positive Genus}, preprint arXiv:1002.3682 [math.PR].}
\nref\Bensimon{see for instance M. Mutz and D. Bensimon, {\it Observation of 
toroidal vesicles}, Phys. Rev. {\bf A 43}, (1990) 4525-4527;
X. Michalet and D. Bensimon, {\it Vesicles of Toroidal Topology: 
Observed Morphology and Shape Transformations}, J. Physique II {\bf 5} (1995)
263-287.}
\nref\TJ{T. Jonsson, {\it Width of handles in two-dimensional quantum gravity},
Phys. Lett. {\bf B 425} (1998) 265-268, arXiv:hep-th/9801150.}
%\nref\FOMAP{J. Bouttier, P. Di Francesco and E. Guitter. {\it Blocked edges 
%on Eulerian maps and mobiles: Application to spanning trees, hard particles 
%and the Ising model}, 	J. Phys. A: Math. Theor. {\bf 40} (2007) 7411-7440, 
%arXiv:math.CO/0702097.}
%\nref\CS{P. Chassaing and G. Schaeffer, {\it Random Planar Lattices and 
%Integrated SuperBrownian Excursion}, 
%Probability Theory and Related Fields {\bf 128(2)} (2004) 161-212, 
%arXiv:math.CO/0205226.}
%\nref\MW{G. Miermont and M. Weill, {\it Radius and profile of random planar
%maps with faces of arbitrary degrees}, Electron. J. Probab. 
%{\bf 13} (2008) 79-106, arXiv:0706.3334 [math.PR].}
%\nref\STATGEOD{J. Bouttier and E. Guitter, {\it Statistics of
%geodesics in large quadrangulations}, J. Phys. A: Math. Theor. {\bf 41} 
%(2008) 145001 (30pp), arXiv:0712.2160 [math-ph].}
%\nref\Mier{G. Miermont, {\it Tessellations of random maps of arbitrary
%genus}, Ann. Sci. \'Ec. Norm. Sup\'er., to appear, arXiv:0712.3688 [math.PR]}

%text
\newsec{Introduction}

The understanding of random maps is a fundamental issue in combinatorics 
and many map enumeration results were obtained over the years, using for
instance recursive decomposition \TUT, matrix integrals [\xref\BIPZ,
\xref\PDF] or bijective methods [\xref\CORV-\xref\MOB]. Recall that a map
of genus $h$ is a proper embedding of a graph in ${\cal S}_h$, the compact
oriented surface of genus $h$ without boundary. By random maps, we mean
generically some statistical ensemble of maps distributed according 
to some particular given law. Of particular interest 
in the scaling limit of large random maps, which converge toward nice universal 
probabilistic objects whose metric properties are only partially understood. 
This limit is especially relevant to physics, where maps are used as discrete 
models for fluctuating surfaces, e.g.\ in the field of biological membranes or 
that of string theory. In physics, maps are often equipped with statistical 
models, such as spins or particles, presenting a large variety of critical 
phenomena. This gives rise to many possible sensible scaling limits of 
continuous surfaces, each defining a particular universality class of maps. 
Their classification and characterization are the aim of the so-called 
two-dimensional quantum gravity [\xref\QGRA,\xref\DGZ], which also has deep 
connections with SLE processes [\xref\Duplantier,\xref\BauerBernard].

The simplest universality class is that of the so-called pure gravity, 
which describes the scaling limit of large maps with prescribed 
face degrees, such as triangulations (maps with faces of degree 3 only) or
quadrangulations (maps with faces of degree 4 only), possibly equipped
with non-critical statistical models. In this universality class, a 
particular attention was payed to the limit of large {\it planar} maps, 
i.e.\ maps with the topology of the two-dimensional sphere, converging to 
the so-called Brownian map [\xref\MARMO,\xref\LEGALL].

Even if its topology remains that of a sphere [\xref\LGP,\xref\MierS], 
the Brownian map presents nevertheless intriguing metric properties such as 
a remarkable phenomenon of confluence of its geodesics \LEGALLGEOD, which 
reveals an underlying tree-like structure. More quantitatively, 
the geometry of the Brownian map was characterized by a number of 
distributions measuring its distance statistics, such as its two-point 
function [\xref\AW-\xref\GEOD], which is the law for the distance 
between two random points in the map and its three-point function \THREEPOINT, 
which is the joint law for the three mutual distances between three points.  
Other refined distributions, measuring e.g.\ the length of ``separating loops" 
or that of the common part of confluent geodesics, were also obtained \LOOP. 
Beside the spherical topology, the case of maps with a single boundary 
was also considered \PSEUDO, with an explicit derivation of the law for 
the distance of a random point to the boundary.

The most advanced results on the distance statistics in maps were obtained in 
the context of planar quadrangulations, using a bijection by 
Schaeffer \SThesis\ which gives a coding for these maps by so-called 
{\it well-labeled trees}, i.e.\ plane trees whose vertices carry integer 
labels with particular 
constraints. This approach turned out to be the most adapted to address
distance-related question since the labels precisely encode some of the
distances in the original map. An extension of the Schaeffer bijection 
was also found by Marcus and Schaeffer [\xref\MS-\xref\CMS], 
which establishes a bijection
between bipartite quadrangulations of arbitrary genus $h$ and so-called
{\it well-labeled $h$-trees}, which are maps of genus $h$ with one face only, 
and whose vertices carry integer labels with the same constraints as
in the planar case. In view of this result, a very natural question is 
therefore that of the distance statistics in maps of arbitrary genus.  
The Marcus-Schaeffer bijection was used very recently \Bett\
to discuss the scaling limit of large bipartite quadrangulations of 
fixed genus, showing in particular the convergence to a limiting object  
with Hausdorff dimension $4$, yet to be characterized.

The purpose of this paper is precisely to give
a quantitative characterization of the distance statistics in large toroidal
maps, i.e.\ maps of genus $1$. Here it is worth mentioning that toroidal
cells or vesicles do exist experimentally \Bensimon\ and some of our 
results might
be relevant to their description. We will in particular derive an explicit 
expression for the two-point function of large toroidal maps, which differs
from that of the planar case. In the case of genus $1$, we may also define a
non-trivial ``one-point function" by measuring the length of the shortest
{\it non-contractible} loop passing via a given point. We give here 
the corresponding limiting probability distribution for large maps. 
These results are 
obtained again in the simplest context of genus $1$ bipartite 
quadrangulations, using the Marcus-Schaeffer bijection above 
with well-labeled $1$-trees.  

The paper is organized as follows: in Section 2, we present our
basic tools which are a number generating functions for well-labeled
(plane) trees with marked vertices, with a particular emphasis on their 
scaling limit. We then recall in Section 3 the Marcus-Schaeffer 
bijection between pointed (i.e.\ with a marked vertex called the origin)
bipartite quadrangulations of genus $h$ and well-labeled $h$-trees. 
As an exercise, we show how to recover the number of genus $1$ 
bipartite quadrangulations from our basic generating functions 
of Section 2. We then compute in Section 4 two different one-point functions 
for large pointed toroidal maps: the probability distribution for the
length of the shortest non-contractible 
loop passing via the origin of the map and that for the length of 
the "second-shortest" loop, which is the shortest among non-contractible loops 
not homotopic to the shortest non-contractible one (or its powers). 
We finally derive in Section 5
the expression for the two-point function of large toroidal maps,
which is the law for the distance from the origin of a random 
point in the map. We gather a few concluding remarks in Section 6.

\newsec{Basic tools: generating functions for well-labeled trees}

In this section, we recall a number of known expressions for generating 
functions of {\it well-labeled trees} and derive a few new ones. As will 
be apparent later, the explicit formulas displayed here will serve as
basic tools to obtain, in the next sections, various distance related 
generating functions in ensembles of large toroidal maps.

By well-labeled tree, we mean a plane tree whose vertices carry integer
labels such that:
\item{(i)} the labels on two vertices adjacent in the tree differ by
at most $1$;
\item{(ii)} the minimum label is 1.
\par
It is useful to also introduce almost well-labeled trees where (ii)
is replaced by the weaker condition:
\item{(ii)'} all labels are larger than or equal to 1.
\par
\noindent

The first generating function of interest is that, $R_\ell\equiv R_\ell(g)$,
of {\it planted} almost well-labeled trees, with a weight $g$ 
per edge, and with root label $\ell$ ($\ell \geq 1$). In other words, the
coefficient of $g^n$ in the expansion of $R_\ell$ as a power series in $g$ 
gives the number of planted almost well-labeled trees with $n$ edges and 
root label $\ell$. This generating function was computed 
in \GEOD, with the result:
\eqn\Rell{\eqalign{& R_\ell=R\, {(1-x^\ell)(1-x^{\ell+3})\over (1-x^{\ell+1})
(1-x^{\ell+2})}\cr 
& {\rm with}\ R={1-\sqrt{1-12g}\over 6g}\ \ {\rm and}\  x+{1\over x}+1=
{1\over g R^2}\ .\cr}}
This particular form is obtained by solving of the recursion relation
\eqn\relRell{R_\ell={1\over 1-g (R_{\ell+1}+R_\ell+R_{\ell-1})}\ ,}
with initial condition $R_0=0$, expressing that an almost well-labeled tree 
with root label $\ell$ is coded by the sequence of its descending subtrees,
themselves almost well-labeled trees with root label $\ell$ or $\ell\pm 1$.

The contribution to $R_\ell$ of large trees is encoded in the behavior of this 
generating function in the vicinity of the critical point $g=1/12$. Setting
\eqn\glim{g={1\over 12}(1-\epsilon^2)}
with $\epsilon \to 0$, a non-trivial scaling limit is reached for $R_\ell$
by letting $\ell$ become large as
\eqn\elltoL{\ell= {L\over \sqrt{\epsilon}}}
with $L$ finite. In this limit, we have the expansion
\eqn\expanRell{\eqalign{& R_\ell=2\left(1-\epsilon\, {\cal F}(L)
+{\cal O}(\epsilon^{3/2}) \right)\cr
&{\rm with}\ {\cal F}(L)=1+{3\over \sinh^2\left(\sqrt{{3\over 2}}L\right)}
\ .\cr}}
The scaling function ${\cal F}$ satisfies the non-linear differential equation 
\eqn\eqforF{{\cal F}''= 3({\cal F}^2-1)}
which is the continuous counterpart of eq.~\relRell, obtained by expanding
this discrete equation at order $\epsilon^2$.

Another basic generating function is that, $X_\ell$ ($\ell \geq 1$), of 
almost well-labeled trees with {\it two distinct} (and distinguished) 
{\it marked vertices carrying the same label} $\ell$ and such that the labels on
the (unique) shortest path in the tree joining these two vertices are all 
larger than or equal to $\ell$. Adding for convenience a trivial contribution 
$1$ to $X_\ell$, we have the recursion relation 
\eqn\relXell{X_{\ell}=1+g R_\ell^2X_\ell(1+g R_{\ell+1}^2 X_{\ell+1})}
obtained by inspecting the first occurrence of a label $\ell$ on the shortest
path joining the two marked vertices. The solution of \relXell\ reads
\eqn\Xell{X_\ell={(1-x^3)(1-x^{\ell+1})^2(1-x^{2\ell+3})\over
(1-x)(1-x^{\ell+3})^2(1-x^{2\ell+1})}}
with $x$ as above. In the scaling limit, we have the expansion
\eqn\expanXell{\eqalign{& X_\ell=3\left(1-\sqrt{\epsilon}\, {\cal C}(L)
+{\cal O}(\epsilon) \right)\cr
&{\rm with}\ {\cal C}(L)=\sqrt{6}\ {2+\cosh(\sqrt{6} L)\over \sinh(\sqrt{6}L)}\cr}}
while eq.~\relXell, expanded at leading order in $\epsilon$, translates into 
\eqn\eqforC{{\cal C}'={\cal C}^2-6{\cal F}\ .}
From the explicit forms of ${\cal F}$ and ${\cal C}$ above, we observe that 
me may write the identity
\eqn\CFrel{{\cal C}=-{{\cal F}''\over {\cal F}'}=-\big(\log (-{\cal F}')\big)'
}
since $\vert {\cal F}'(L)\vert =-{\cal F}'(L)$ for real $L$.
In this form, the relation \eqforC\ is a direct consequence of the 
relation \eqforF.

A third useful generating function is that, ${\tilde X}_{\ell_1,\ell_2}$ 
($\ell_1 > \ell_2 \geq 1$), of almost well-labeled trees with 
{\it two distinct} (and distinguished) {\it marked vertices carrying 
respective labels $\ell_1$ and $\ell_2$}, and such that labels on the 
(unique) shortest path in the tree joining them are all strictly larger than 
$\ell_2$ (except of course
at the endpoint of the path with label $\ell_2$). By cutting this
shortest path at the first occurrence of a label $\ell_1-1, \ell_1-2, \cdots$, 
we may write 
\eqn\tildeX{\eqalign{{\tilde X}_{\ell_1,\ell_2}
&=\prod_{\ell=\ell_2}^{\ell_1-1} g R_\ell R_{\ell+1} X_{\ell+1}\cr 
&= x^{\ell_1-\ell_2} {(1\!-x^{\ell_2})(1\!-x^{\ell_2+1})
(1\!-x^{\ell_2+2})(1\!-x^{\ell_2+3}) (1\!-x^{2\ell_1+3})\over (1\!-x^{\ell_1})
(1\!-x^{\ell_1+1}) (1\!-x^{\ell_1+2})(1\!-x^{\ell_1+3})(1\!-x^{2\ell_2+3})}
\ .\cr}}
This last expression is extended to the case $\ell_1=\ell_2$ by adopting 
the convention that ${\tilde X}_{\ell_2,\ell_2}=1$. 
In the scaling limit, setting $\ell_1=L_1/\sqrt{\epsilon}$ and 
$\ell_2= L_2/\sqrt{\epsilon}$, we have the expansion 
\eqn\expantildeX{\eqalign{&{\tilde X}_{\ell_1,\ell_2}= 
{\tilde {\cal C}}(L_1,L_2)+
{\cal O}(\sqrt{\epsilon})\cr &{\rm with}\ {\tilde {\cal C}}(L_1,L_2)= {
\cosh\left(\sqrt{{3\over 2}}L_1\right)\sinh^3\left(\sqrt{{3\over 2}}L_2
\right)\over
\cosh\left(\sqrt{{3\over 2}}L_2\right)\sinh^3\left(\sqrt{{3\over 2}}L_1
\right)}\ .\cr}}
Note that this expression can be obtained alternatively by looking 
at the continuous counterpart of the product formula in \tildeX. Indeed, 
using $g R_\ell R_{\ell+1} X_{\ell+1} \sim (1/12)\times 2^2 \times 3\, 
\left(1-\sqrt{\epsilon}\,{\cal C}(L)\right)$, we may write directly
\eqn\directC{{\tilde {\cal C}}(L_1,L_2)= \exp\left(-\int_{L_2}^{L_1}
{\cal C}(L)\, dL\right) = \exp\left(-\big[-\log (-{\cal F}')
\big]_{L_2}^{L_1}
\right)= {{\cal F}'(L_1)\over {\cal F}'(L_2)}}
which matches \expantildeX.

The final important quantity is what we shall call the ``propagator",
which is the generating function $K_{\ell_1,\ell_2}$
($\ell_1,\ell_2\geq 1$) of almost well-labeled trees with {\it two distinct}
(and distinguished) {\it marked vertices with respective labels $\ell_1$
and $\ell_2$}, and with no additional condition on the labels in-between
(apart of course from the general conditions (i) and (ii)' of almost
well-labeled trees). This generating function reads
\eqn\Krel{K_{\ell_1,\ell_2}= 
-\delta_{\ell_1,\ell_2}+\sum_{\ell=1}^{\min(\ell_1,\ell_2)}
{\tilde X}_{\ell_1,\ell} {\tilde X}_{\ell_2,\ell} X_\ell}
as obtained by summing over the minimum label $\ell$ on the path
joining the two marked vertices. Note that we insist here on
the two marked vertices being distinct, with a path of non-zero length 
in-between and that, if $\ell_1=\ell_2$, we have to subtract an 
undesired contribution $1$ in the sum. The generating function
$K_{\ell_1,\ell_2}$ is solution of the equation
\eqn\recurK{\eqalign{K_{\ell_1,\ell_2}=g R_{\ell_1}\Big(R_{\ell_1+1}
(\delta_{\ell_1+1,\ell_2}+ K_{\ell_1+1,\ell_2}) &
+R_{\ell_1}(\delta_{\ell_1,\ell_2}+K_{\ell_1,\ell_2})\cr &
+R_{\ell_1-1}(\delta_{\ell_1-1,\ell_2}+K_{\ell_1-1,\ell_2})\Big)\cr}}
with $K_{0,\ell_2}=0$, obtained by looking at the label 
$\ell_1, \ell_1\pm 1$ of the vertex 
adjacent to the first marked vertex on the path joining the two marked 
vertices.
We have no compact formula for $K_{\ell_1,\ell_2}$ at the discrete level
but, in the scaling limit, eq. \Krel\ translates into the expansion
\eqn\expandK{\eqalign{&
K_{\ell_1,\ell_2}={\rho(L_1,L_2) \over  \sqrt{\epsilon}}
+{\cal O}(1)\cr &{\rm with}\ {\rho}(L_1,L_2)=
\int_{0}^{\min(L_1,L_2)} \kern-30pt 3\, {\tilde {\cal C}}(L_1,L){\tilde {\cal C}}(L_2,L)
\ dL\cr}}
with a simple factor $3$ for the limit of $X_\ell$. Using \directC, we may 
now compute the integral and write a compact expression for $\rho(L_1,L_2)$:
\eqn\compactrho{\eqalign{& \rho(L_1,L_2)= 
{\cal F}'(L_1){\cal F}'(L_2)\times {\cal H}\big(\min(L_1,L_2)
\big)\cr
&{\rm where}\ {\cal H}(L)= {1\over 2^6 3^3} 
\left(\!180 L\!+\!\sqrt{6} \left(\!\sinh \left(2 \sqrt{6} L\right)\!
-\!16 \sinh \left(\sqrt{6} L\right)\!-\!32 \tanh \left(\sqrt{{3\over 2}}
L\right)\!\right)\!\right). \cr}}
Note that the somewhat involved function ${\cal H}(L)$ is simply 
characterized by 
\eqn\carf{{\cal H}'(L)={3\over \left({\cal F}'(L)\right)^2}}
with ${\cal H}(0)=0$. It is a straightforward exercise to check that the 
continuous propagator $\rho(L_1,L_2)$ above is solution of the equation
\eqn\eqforrho{{\partial^2 \ \over \partial L_1^2}\rho(L_1,L_2)= 6 {\cal F}(L_1)
\rho(L_1,L_2)-3\, \delta(L_1-L_2)}
which is the continuous counterpart of \recurK.
Up to a (somewhat arbitrary) global normalization of the coordinate $L$, 
this is precisely the equation satisfied by the two-point function of a 
polymer chain embedded in one dimension (with coordinate $L$) subject to
a potential $V(L)\propto {\cal F}(L)-1 \propto 1/\sinh^2(\sqrt{{3/2}}L)$. 
Heuristically, this polymer corresponds to the (unique) path between the 
two marked points in the tree and its configuration in one-dimensional space
simply reproduces the sequence of labels along this path. From condition (ii)',
the presence of the other attached subtrees exerts on the polymer an
effective repulsion from the position $L=0$ encoded in the potential $V(L)$. 

When $\ell_1\to \infty$, $K_{\ell_1,\ell_2}\to 0$ unless we also send
$\ell_2\to \infty$, keeping $p\equiv \ell_2-\ell_1$ finite.
Defining 
\eqn\trinv{k_p\equiv \lim_{\ell_1\to \infty}K_{\ell_1,\ell_1+p}\ ,}
for an arbitrary integer $p$, this quantity satisfies the equation
\eqn\recurk{k_p=g\, R^2 (k_{p-1}+k_p+k_{p+1}+\delta_{p,1}+\delta_{p,0}+
\delta_{p,-1})}
and we find explicitly
\eqn\valkp{k_p={1-x^3 \over (1-x)(1-x^2)}\, x^{\vert p\vert} -\delta_{p,0}}
with $x$ as in \Rell.
In the scaling limit, and for large values of $L_1$ and $L_2$, we have the 
corresponding limiting behavior
\eqn\limitrho{\rho(L_1,L_2)\sim {\sqrt{6}\over 4} \exp(-\sqrt{6}\, \vert
L_1-L_2\vert)}
where we recognize the usual (translation invariant) two-point function
for a polymer in one dimension without potential.

The above quantities $X_\ell$, ${\tilde X}_{\ell_1,\ell_2}$ and 
$K_{\ell_1,\ell_2}$ may alternatively be viewed as generating functions 
for {\it chains} of almost well-labeled trees, with endpoints carrying 
prescribed labels. In the following, we will consider various generating 
functions for more involved well-labeled structures coding for toroidal maps. 
At the discrete level, those are naturally formed of a number of chains 
as above, attached by their endpoints. The knowledge of the scaling forms
${\cal C}(L)$, ${\tilde {\cal C}}(L_1, L_2)$ and $\rho(L_1,L_2)$ for the 
chain generating functions will be sufficient to derive explicit expressions
for the scaling form of these more involved toroidal generating functions. 
This eventually will translate into explicit universal scaling 
functions characterizing the distance statistics of large toroidal maps.  

\newsec{The Marcus-Schaeffer bijection for bipartite quadrangulations
of genus $1$}

Our approach on the distance statistics of large toroidal maps relies on 
the bijection by Marcus and Schaeffer between, one the one hand, 
{\it pointed bipartite quadrangulations of fixed genus $h$} and on the 
other hand, {\it well-labeled $h$-trees}. By pointed bipartite quadrangulation,
we mean a map whose faces all have degree $4$ (quadrangulation), whose 
vertices can be colored in black and white in such a way that adjacent 
vertices have a different color (bipartite) and with a marked vertex 
called the origin (pointed). It was shown by Marcus and Schaeffer \MS\ that 
any such map of genus $h$ is bijectively coded by a well-labeled $h$-tree, 
i.e.\ a map of genus $h$ with exactly one face ($h$-tree) and
whose vertices carry integer labels subject to the two conditions (i) and (ii) 
of section 2 (well-labeled). As before, it is convenient to also introduce
the notion of almost well-labeled $h$-tree where condition (ii) is replaced
by the weaker condition (ii)'.

Note that a quadrangulation with $n$ faces and genus $h$ has $2n$ edges and
$n+2-2h$ vertices. The $n$ faces of the quadrangulation are in one-to-one
correspondence with the edges of the well-labeled $h$-tree, which therefore 
has $n$ edges and $n+1-2h$ vertices. These $n+1-2h$ vertices are in one-to-one 
correspondence with the $(n+2-2h)-1$ vertices of the
quadrangulation other than its origin, and the label of a vertex in
the $h$-tree is nothing but the {\it distance from the associated vertex 
to the origin} in the quadrangulation. Finally, the $2n$ corners of the $h$-tree
(i.e.\ the angular sectors between consecutive edges around a vertex)
are in one-to-one correspondence with the $2n$ edges of the
quadrangulation. More precisely, the corners with label $\ell$ (i.e.\ 
around a vertex with label $\ell$) in the 
$h$-tree are in one-to-one correspondence with the edges of type 
$(\ell-1)\to\ell$ in the quadrangulation, i.e.\
the edges connecting a vertex at distance $(\ell-1)$ from the origin
to a vertex at distance $\ell$. 

It is useful to recall how to recover the quadrangulation from its 
well-labeled $h$-tree coding. For the vertices of the quadrangulation,
we take all the vertices of the $h$-tree plus an extra vertex added inside 
its unique face. This vertex will be the origin of the (pointed)
quadrangulation. The edges of the quadrangulation are obtained by {\it linking 
each corner with label $\ell$ in
the $h$-tree to its successor}, which is the added vertex for $\ell=1$
and, for $\ell>1$, the first corner with label $\ell-1$ encountered, 
say counterclockwise around the unique face of the $h$-tree. These links
can be drawn without mutual crossings inside the face of the $h$-tree and, 
by construction, they do not intersect the original edges of this $h$-tree. 
Note that the sequence
of links between the successive successors of a given corner with label $\ell$ 
provides a particular geodesic (i.e.\ shortest) path of length $\ell$ in the 
quadrangulation, leading from the vertex underlying this corner to the origin
vertex. After drawing the 
quadrangulation edges, we may erase all the original 
edges of the $h$-tree as well as all the labels. 
Note that the obtained quadrangulation is automatically bipartite. 
\fig{The two possible backbones for $1$-trees: (a) a generic backbone
with two vertices of degree $3$ carrying positive labels $\ell_1$ and $\ell_2$
linked by three edges and (b) a degenerate backbone with one vertex of
degree $4$ with positive label $\ell$, linked to itself by two 
edges.}{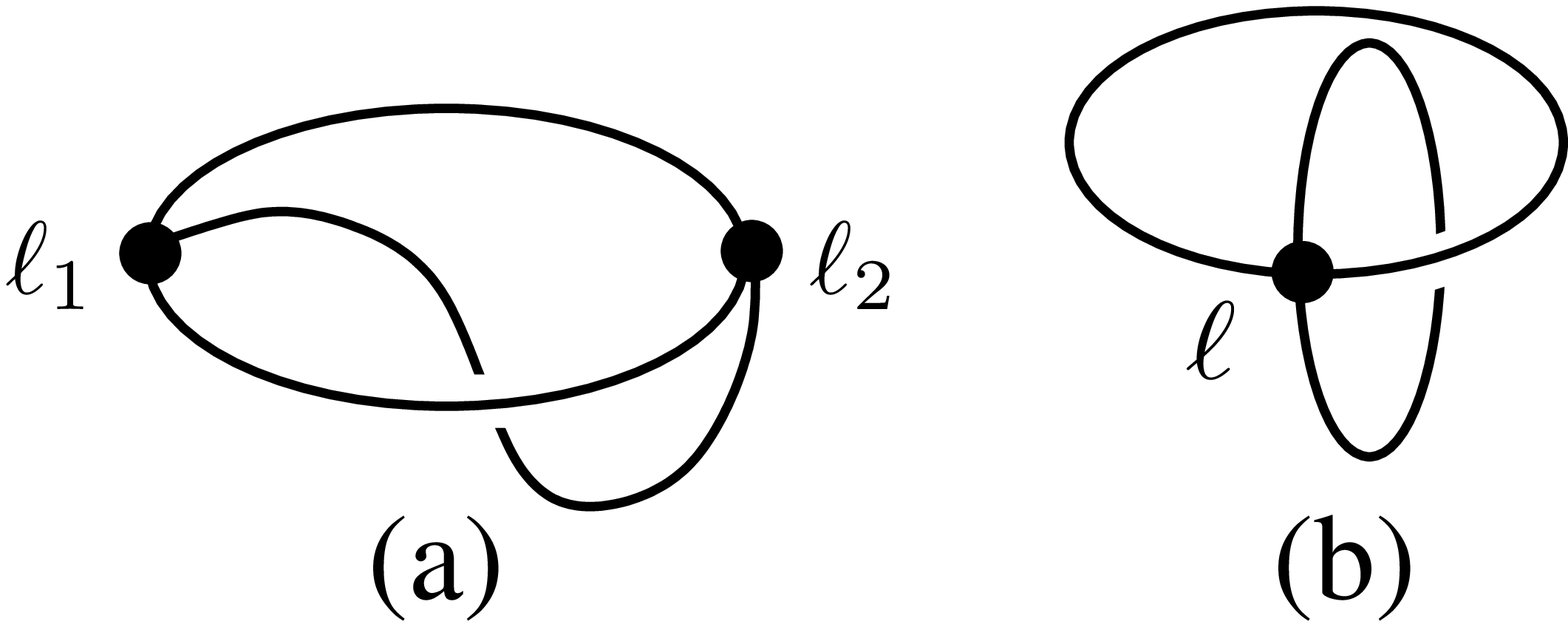}{8.cm}
\figlabel\backbones
In the spherical case $h=0$, the well-labeled $0$-trees are nothing
but the well-labeled trees of Section 2, so that our generating functions
above translate into generating functions for pointed quadrangulations
(note that we need not specify that the quadrangulation is bipartite 
since this is automatic for $h=0$). For instance, choosing a corner with 
label $\ell$ in a well-labeled tree amounts to picking an edge at distance
$\ell$ from the origin (i.e.\ an edge of type $(\ell-1)\to \ell$) in 
the associated quadrangulation. The generating function for pointed planar
quadrangulations with a marked edge at distance $\ell$ is therefore
that of planted well-labeled trees with root label $\ell$. It is given by 
$R_\ell-R_{\ell-1}$ since we have to subtract from $R_\ell$ those 
configurations with minimum label larger than or equal to $2$. These 
undesired configurations correspond exactly to almost well-labeled 
configurations having all their labels shifted by $1$, hence they are
counted by $R_{\ell-1}$. We will use the same shift argument in the
following to transform generating functions for almost well-labeled 
objects into generating functions for their well-labeled analogs. 
For instance, the generating function for pointed planar quadrangulations 
with two marked distinct (and distinguished) vertices at distance 
$\ell_1$ and $\ell_2$ from the origin is given by 
$K_{\ell_1,\ell_2}-K_{\ell_1-1,\ell_2-1}$
in terms of the propagator. 

In this paper, we focus on the case $h=1$ of toroidal topology. Well-labeled
and almost well-labeled $1$-trees may be classified according to their 
{\it backbone} obtained as follows: we first delete recursively each vertex 
with degree $1$ and its incident edge until no degree $1$ vertex is left
in the $1$-tree.
The resulting object is an almost well-labeled $1$-tree whose vertices
all have degree larger than or equal to $2$ by construction. We
call this $1$-tree the {\it skeleton} of the original $1$-tree. We may 
then erase each vertex of degree $2$ in the skeleton and concatenate the two 
incident edges, leading to a labeled $1$-tree with vertices of degree larger 
than or equal to $3$ and with arbitrary positive vertex labels. We call
this $1$-tree the {\it backbone} of the original $1$-tree. From the Euler 
relation for maps, it is easily seen that only two types of backbones 
are possible:
\item{(a)}{generic backbones, made of two $3$-valent vertices 
with labels $\ell_1\geq \ell_2\geq 1$, linked by three edges
(see figure \backbones-(a));}
\item{(b)}{degenerate backbones, made of a unique $4$-valent vertex
with label $\ell\geq 1$ linked to itself by two edges 
(see figure \backbones-(b)).}
\par
\noindent To obtain all the almost well-labeled $1$-trees leading to a 
generic backbone, we simply have to replace each of the three edges of this 
backbone by an arbitrary non-empty chain of almost well-labeled trees with 
endpoints labeled by $\ell_1$ and $\ell_2$, as counted by the propagator
$K_{\ell_1,\ell_2}$.
The corresponding generating function reads therefore
\eqn\contone{
\eqalign{
&{1\over 3} K_{\ell_1,\ell_2}^3 \quad {\rm if}\ \ell_1>\ell_2\cr
&{1\over 6} K_{\ell_1,\ell_1}^3 \quad {\rm if}\ \ell_1=\ell_2\ .\cr }}
Here a factor $1/3$ is necessary to avoid over-counting as, in the sum, we
recover $3$ times the same configuration upon permuting cyclically
the three edges of the skeleton. Similarly, an extra $1/2$ factor is necessary
to balance the possibility of recovering the same configuration by 
exchanging the two vertices of degree $3$ if they have the same label.
This simple over-counting argument does not hold for configurations 
presenting a symmetry and these configurations are therefore enumerated 
in \contone\ with an inverse symmetry factor, which is a customary statistics. 
Note that the symmetric configurations are expected to 
become negligible for maps of large size $n$ so that the symmetry factors
have no effect on the scaling limit that we will discuss.

If we wish to keep only the well-labeled $1$-trees, we have to remove those 
configurations
for which the minimum label is larger than or equal to $2$ and, using
the same shift argument as above, we now get a generating function
\eqn\conttwo{
\eqalign{
&{1\over 3} \left(K_{\ell_1,\ell_2}^3 -K_{\ell_1-1,\ell_2-1}^3\right)
\quad {\rm if}\ \ell_1>\ell_2\cr
&{1\over 6} \left(K_{\ell_1,\ell_1}^3 -K_{\ell_1-1,\ell_1-1}^3\right)
\quad {\rm if}\ \ell_1=\ell_2\cr }}
with the convention $K_{\ell_1,0}=K_{0,0}=0$. 

Summing first over the smallest label $\ell_2$, then over the largest one
$\ell_1$, we get the total contribution $W_1$ of all $1$-trees having a 
generic backbone 
\eqn\totalcont{\eqalign{W_1&= \sum_{\ell_1=1}^\infty \left\{{1\over 6}
\left(K_{\ell_1,\ell_1}^3 -K_{\ell_1-1,\ell_1-1}^3\right)
+{1\over 3}\sum_{\ell_2=1}^{\ell_1-1} 
\left(K_{\ell_1,\ell_2}^3 -K_{\ell_1-1,\ell_2-1}^3\right)\right\}\cr
&=\sum_{\ell_1=1}^\infty\left\{
\left({1\over 6}K_{\ell_1,\ell_1}^3 +{1\over 3}\sum_{\ell_2=1}^{\ell_1-1}
K_{\ell_1,\ell_2}^3\right)-
\left({1\over 6}K_{\ell_1-1,\ell_1-1}^3 +{1\over 3}\sum_{\ell_2=1}^{\ell_1-2}
K_{\ell_1-1,\ell_2}^3\right)\right\}\cr
&= \lim_{\ell_1\to \infty} \left\{{1\over 6}K_{\ell_1,\ell_1}^3+{1\over 3}
\sum_{\ell_2<\ell_1} K_{\ell_1,\ell_2}^3\right\}\cr
&= {1\over 6} k_0^3+ {1\over 3}\sum_{p<0}k_p^3 = {x^3(1+2x+2x^2-2x^3)\over
2(1-x)^4(1+x)^2}\ .\cr}}
Similarly, the total contribution $W_2$ of all well-labeled 
$1$-trees having a degenerate backbone is
\eqn\totalcontdeg{\eqalign{W_2&= \sum_{\ell_1=1}^\infty {1\over 4}
\left(K_{\ell_1,\ell_1}^2 -K_{\ell_1-1,\ell_1-1}^2\right)\cr
& = \lim_{\ell_1\to \infty} {1\over 4}K_{\ell_1,\ell_1}^2\cr
& = {1\over 4} k_0^2 = {x^2(1+2x)^2 \over 4(1-x)^2(1+x)^2}\ .\cr}}
Summing \totalcont\ and \totalcontdeg, we get the generating function
for well-labeled $1$-trees which, from the Marcus-Schaeffer bijection is also
that, $Q^{(1)}_\bullet(g)$ of pointed bipartite quadrangulations of genus 
$1$
\eqn\Qone{Q^{(1)}_\bullet= W_1+W_2={x^2(1+4x+x^2)\over 4(1-x)^4(1+x)^2}\ .}
Note that the symmetry factor 
of a well-labeled $1$-tree is also that of the associated pointed 
quadrangulation so that symmetric pointed quadrangulations are 
counted in $Q^{(1)}_\bullet$ with their usual inverse symmetry factor.
It is more customary to consider {\it rooted} maps, i.e.\ maps
with a marked oriented edge, as they do not involve symmetry factors. 
The generating function for rooted bipartite quadrangulations of genus $1$ 
is simply
\eqn\rootedQ{Q^{(1)}_\to= 4 Q^{(1)}_\bullet}
since there are exactly twice as many edges as vertices in a genus $1$
quadrangulation, each coming with two orientations. All the enumeration
formulas above are consistent with those found in ref.~\CMS.

If we are interested only in large maps, we can use the continuous analogs
of \totalcont\ and \totalcontdeg\ giving the leading singularity of 
$W_1$ and $W_2$, namely
\eqn\Wcount{\eqalign{W_1&\sim {1\over \epsilon^2} \lim_{L_1\to \infty}
{1\over 3}\int_0^{L_1}dL_2\, \rho^3(L_1,L_2) \cr & = {1\over 3 \epsilon^2}
\int_{-\infty}^0 dP\,  
\left({\sqrt{6}\over 4}\right)^3 \exp(-3 \sqrt{6}\, \vert P\vert )= 
{1\over 96 \epsilon^2} \cr 
W_2 &\sim {1\over \epsilon} \lim_{L_1\to \infty} {1\over 4} \rho^2(L_1,L_1)
= {1\over 4 \epsilon} \left({\sqrt{6}\over 4}\right)^2 = {3\over 32 \epsilon}
\cr}}
which can also be directly read off \totalcont\ and \totalcontdeg\ with
$x\sim \exp(-\sqrt{6\, \epsilon})$. Note that the degenerate case
is less singular than the generic case so that the dominant singularity
of $Q^{(1)}_\bullet$ or $Q^{(1)}_\to$ comes from $W_1$ only. 

Alternatively, we deduce from \Wcount\ the large $n$ behavior of the number 
of well-labeled $1$-trees with $n$ edges having a generic or degenerate 
backbone
\eqn\Wonegn{W_1\vert_{g^n}\sim {12^n\over 96\ }\ , \qquad 
W_2\vert_{g^n}\sim {3\ 12^n \over 32\sqrt{\pi n}}\ .}
At large $n$, the dominant contribution to $Q^{(1)}_\bullet$ or $Q^{(1)}_\to$
comes from configurations with a generic backbone only and we recover the 
known asymptotics \CMS
\eqn\asymQ{Q^{(1)}_\to\vert_{g^n}=4 Q^{(1)}_\bullet\vert_g^n 
\sim {12^n \over 24\ }} 
for the number of rooted bipartite quadrangulations of genus $1$ with 
$n$ faces.

\newsec{One-point functions for large toroidal maps}
\fig{A non-contractible loop (thick red) drawn on a toroidal quadrangulation
must intersect the skeleton of the associated $1$-tree at one of its vertices.
}{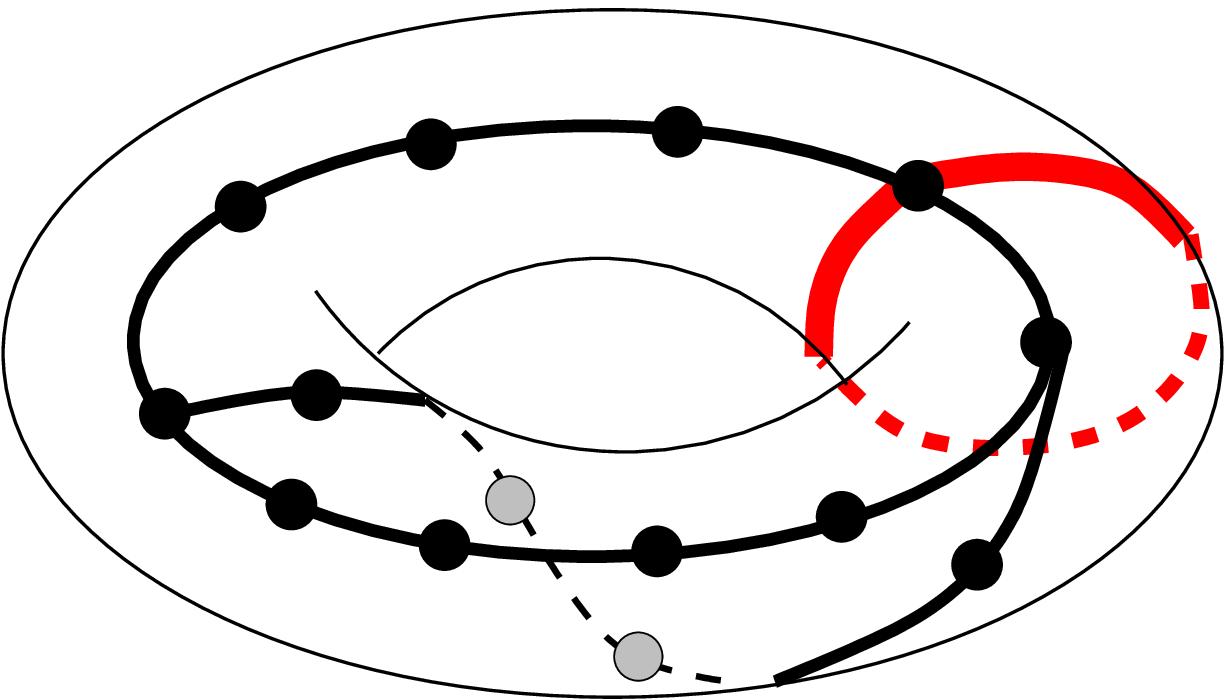}{6.cm}
\figlabel\noncont
For maps with a toroidal topology, we may define several interesting
distance-dependent one-point functions. Starting with a pointed
map, we may consider as a``self-distance" of the origin vertex the length 
of any shortest {\it non-contractible} 
loop in the map containing this origin. In the case of bipartite 
quadrangulations, we denote this length by $2\ell$ as it is necessarily 
even. We may now use the coding of previous section and note that any 
non-contractible loop in the quadrangulation must intersect
the skeleton of the associated $1$-tree. Now the edges of the quadrangulation 
lie strictly inside the unique face of the $1$-tree and do not cross its edges
so that the intersection with the skeleton necessarily takes place 
at one of the skeleton vertices, say with label $\ell'$
(see fig.~\noncont\ for an illustration). Since $\ell'$ is the distance in 
the quadrangulation from this vertex to the origin, the length of the loop is 
necessarily larger than or equal to $2\ell'$ so that the half-length $\ell$ 
of a shortest non-contractible loop is larger than or equal to the 
minimum label on the skeleton of the $1$-tree. Picking now a vertex with 
minimal label $\ell_{\rm min}$ on this skeleton, this vertex 
has a degree larger than or equal to $2$ by construction and we may consider
two distinct corners around it. The two sequences of links between the
successive successors of these corners define two paths of lengths 
$\ell_{\rm min}$ leading to the origin, which {\it do not intersect 
the skeleton} and whose concatenation therefore creates a non-contractible 
loop of length $2\ell_{\rm min}$. We deduce that 
\eqn\llmin{\ell=\ell_{\rm min}({\rm Sk.})}
where $\ell_{\rm min}({\rm Sk.})$ is the minimum label on the skeleton of 
the $1$-tree coding for the pointed quadrangulation at hand. 

In the scaling limit, as already noticed, it is sufficient to consider the
contribution of configurations leading to a generic backbone and we 
denote 
by $\ell_1=L_1/\sqrt{\epsilon}$ and $\ell_2=L_2/\sqrt{\epsilon}$ the 
labels of its two vertices of degree $3$, with $\ell_1\geq \ell_2$, and 
by $m_1=M_1/\sqrt{\epsilon}$, $m_2=M_2/\sqrt{\epsilon}$ and 
$m_3=M_3/\sqrt{\epsilon}$ the respective minimal labels on the three chains 
linking these vertices in the skeleton, with $m_1\geq m_2 \geq m_3$ so
that $\ell_{\rm min}({\rm Sk.})=m_3$. With these notations, the generating
function for almost well-labeled $1$-tree whose skeleton has a minimal
label $\ell_{\rm min}({\rm Sk.})=\ell=L \sqrt{\epsilon}$ behaves 
when $\epsilon\to 0$ as
\eqn\gfnc{\eqalign{& \! {2\over \epsilon^2} 
\int_{M_3}^\infty \!dM_2\int_{M_2}^\infty \!dM_1
\int_{M_1}^\infty \!dL_2 \int_{L_2}^\infty \!dL_1  \ 
3\, {\tilde {\cal C}}(L_1,M_1) {\tilde {\cal C}}(L_2,M_1)
\times 3\, {\tilde {\cal C}}(L_1,M_2) {\tilde {\cal C}}(L_2,M_2)\cr
& \hskip 8.cm \times 3\, {\tilde {\cal C}}(L_1,M_3) {\tilde {\cal C}}(L_2,M_3)\Big\vert_{M_3=L}
\cr
& ={2 \over \epsilon^2}{3\over {\cal F}'(L)^2}
\int_{L}^\infty \!dM_2 {3\over {\cal F}'(M_2)^2} 
\int_{M_2}^\infty \!dM_1 {3\over {\cal F}'(M_1)^2} 
\int_{M_1}^\infty \!dL_2 {\cal F}'(L_2)^3  
\int_{L_2}^\infty \!dL_1 {\cal F}'(L_1)^3
\cr
& = {{\cal I}(L)\over \epsilon^2} \quad \quad {\rm with}\quad  
{\cal I}(L)={1\over 96}
\tanh^4\left(\sqrt{3\over 2}L\right) \ .\cr}
}
Note, as in \expandK, the factors $3$ for the limit of each of the discrete
terms $X_{m_1}$, $X_{m_2}$ and $X_{m_3}$. 
Note also the factor $2$ accounting for the two possible order of
appearance of the minima $m_1$, $m_2$ and $m_3$ when turning around  
say the vertex with larger label $\ell_1$. 

To return to well-labeled configurations, we note that, as
an alternative to the shift argument used in Section 3, we may 
equivalently use the
property that almost well-labeled $1$-trees whose skeleton has a minimal 
label $\ell$ are in one-to-one correspondence with well-labeled $1$-trees whose
skeleton has a minimal label smaller than or equal to $\ell$.
This correspondence is obtained by now shifting all labels in
the almost well-labeled $1$-tree by a non-negative integer so that 
its global minimum label (in the whole $1$-tree) becomes $1$.
From the Marcus-Schaeffer bijection, we deduce that \gfnc\ may alternatively
be viewed as the
$\epsilon\to 0$ leading behavior of the {\it generating function for
pointed bipartite quadrangulations with a shortest non-contractible
loop passing via the origin of (rescaled) length smaller than $2L$}.
\fig{Plots of the cumulative probability distribution $\sigma(r)$ and the 
associated
probability density $\sigma'(r)$ for the rescaled half-length $r$ of
the shortest non-contractible loop passing via the origin of
a large pointed toroidal quadrangulation.}{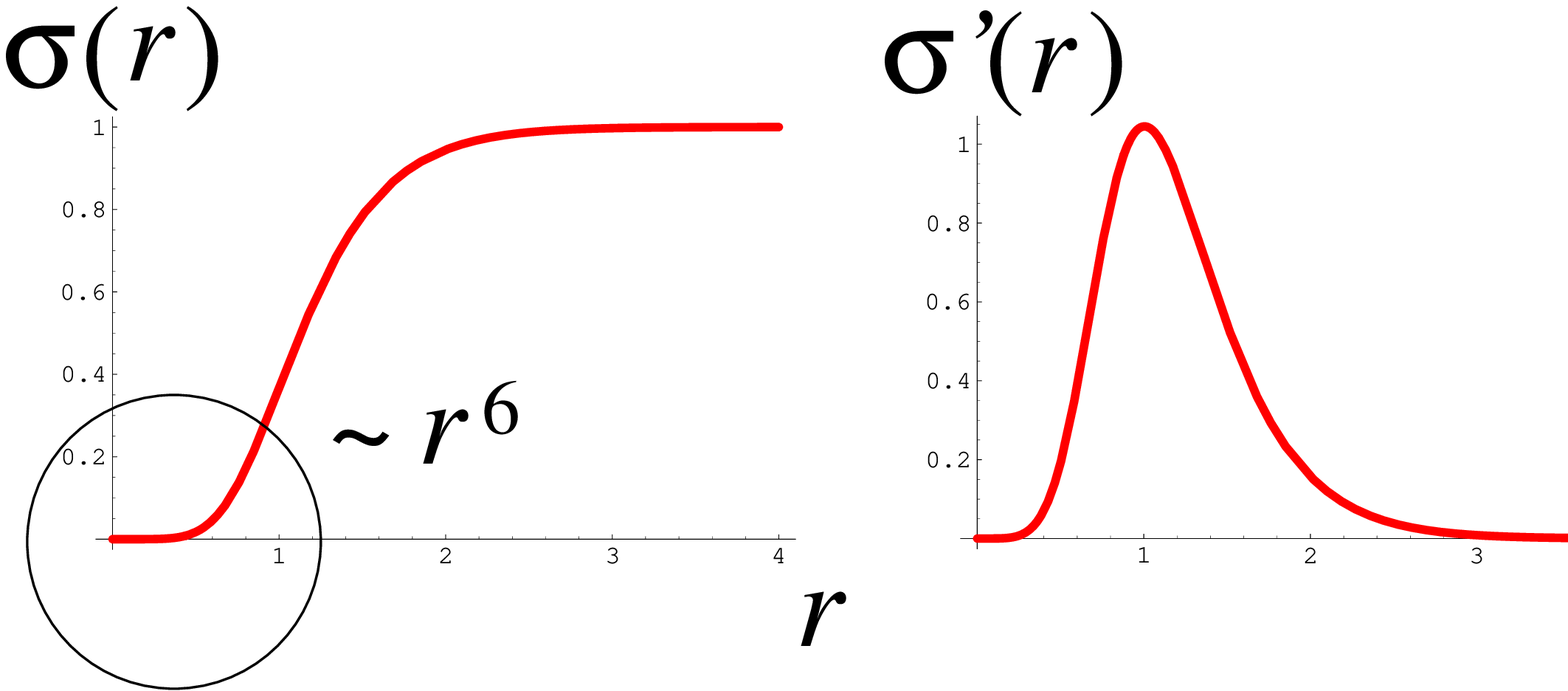}{12.cm}
\figlabel\sigmaplots
From our one-point toroidal scaling function ${\cal I}(L)$, 
we may extract a (cumulative)
{\it probability distribution} in the ensemble of bipartite quadrangulations 
of genus $1$ with fixed size $n$ ($=$ number of faces), in the limit of 
large $n$. 
At the discrete level, fixing the size amounts to picking the coefficient 
$g^n$ in the generating function at hand, which can be done by a contour 
integral in $g$. In the limit of large $n$, this contour integral translates 
into an integral over a real variable $\xi$ upon setting
(see \GEOD\ for details)
\eqn\saln{g={1\over 12}\left(1+{\xi^2\over n}\right)\ .}
A sensible scaling limit is now obtained by rescaling $\ell$ as
\eqn\scalnL{\ell=r\, n^{1/4}}
so that we can use in practice our continuous expressions above with 
$\epsilon=-{\rm i}\xi /\sqrt{n}$ and $L=\sqrt{-{\rm i}\xi}\, r$. 
Normalizing by $Q^{(1)}_\bullet\vert g^n$, we get the {\it probability}
\eqn\sigmalaw{\eqalign{\sigma(r)& ={96\over \pi}\int_0^{\infty}
d\xi\, {1\over -{\rm i}\xi} \exp(-\xi^2) 
\left({\cal I}(\sqrt{-{\rm i}\xi}r)-{\cal I}(\sqrt{{\rm i}\xi}r)\right)\cr
&= {8\over \pi} \int_0^\infty {d\xi\over \xi} \exp(-\xi^2)
{\sin(\sqrt{3\xi}r)\sinh^3(\sqrt{3\xi}r)-\sinh(\sqrt{3\xi}r)
\sin^3(\sqrt{3\xi}r)\over (\cosh(\sqrt{3\xi}r)-\cos(\sqrt{3\xi}r))^4} \cr}}
{\it that the shortest non-contractible loop passing via the 
origin has a length smaller than $2 r n^{1/4}$ in the ensemble of pointed 
bipartite quadrangulations of genus $1$ with fixed size $n$}, 
in the limit $n\to\infty$. The cumulative probability distribution $\sigma(r)$ 
and the associated probability density $\sigma'(r)$ are plotted 
in fig.~\sigmaplots.
For small $r$, we have the expansion
\eqn\expansigma{\sigma(r)={9 r^6\over 4 \sqrt{\pi}}-{1431 r^{10}\over 280 
\sqrt{\pi}}+{\cal O}(r^{14})\ .}
It is worth mentioning that, in the continuous limit, the shortest 
non-contractible loop is expected to be unique, i.e.\ two shortest loops
at the discrete level remain at a distance negligible with respect
to $n^{1/4}$ at large $n$. Moreover, as for geodesic paths in spherical
maps, we expect a confluence phenomenon with the shortest loop made 
of an open part of (rescaled) length $2r-2\delta$ and
a common part of length $\delta$ traveled back and forth in the vicinity
of the origin. The joint law for $\delta$ and $r$ could be obtained in
principle by a slight refinement of the above analysis.
\fig{Given a first non-contractible loop (thick red) intersecting only one of 
the three chains forming the skeleton, any other non-contractible loop 
non-homotopic to
this first loop or to its powers (thick blue) must intersect (at least) 
one of the two other chains in the skeleton (possibly at an
endpoint of the chain).}{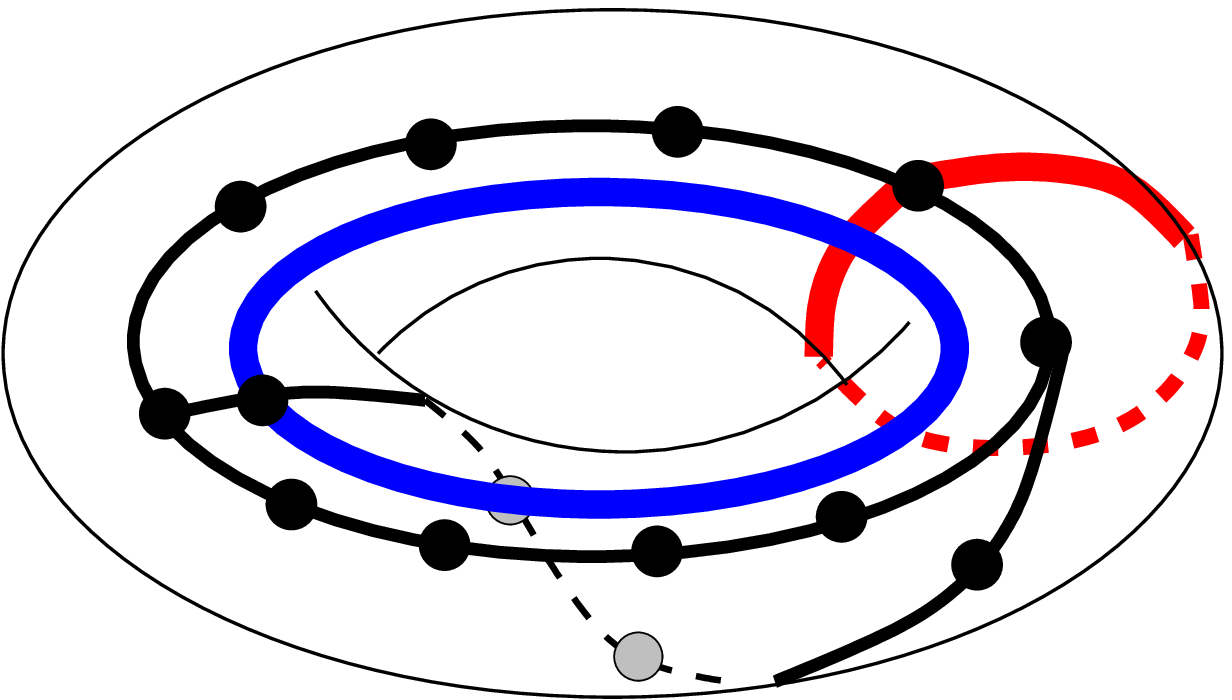}{6.cm}
\figlabel\nonconttwo
In the same spirit, for a pointed map of genus $1$, we may first consider
a shortest non-contractible loop as above and now study the length of any 
shortest loop among those non-contractible loops {\it not homotopic to the 
shortest loop} or to any of its powers. A similar quantity was introduced
in \TJ\ an analyzed using heuristic scaling arguments.
At the discrete level, this notion depends on the particular choice of the
first shortest loop but we expect that this dependence is wiped out
in the continuous limit in which the first shortest loop is
essentially unique. We may as before assume that the well-labeled 
$1$-tree coding for the quadrangulation has a generic backbone, i.e.\ has a
skeleton with two $3$-valent vertices linked by three chains. 
We may moreover assume that the minimum label on this skeleton is 
reached for a bivalent vertex,
i.e.\ strictly inside one of the three chains. Indeed, all the other 
situations (degenerate backbone or minimal label
at one of the two $3$-valent vertices of the skeleton) correspond to 
degenerate cases which give sub-dominant contributions in the continuous limit.
We now take for the first loop the concatenation of the two sequences of links
between the successive successors of the two corners at the above bivalent 
vertex with minimal label. A non-contractible loop not homotopic to this loop 
or to its 
powers must necessarily intersect the skeleton at one of the two other chains
so that its length is larger than twice the label of any vertex of the 
skeleton minus the first chain (see fig.~\nonconttwo\ for an illustration). 
Taking a vertex with minimal label $\ell_{\rm min}$ in this set, 
the two sequences of successors of two corners at this vertex, once
concatenated, create a suitable loop of length $2\ell_{\rm min}$. 

To summarize, denoting as before by $L_1$ and $L_2$ the (rescaled) labels of 
the $3$-valent vertices of the skeleton, with $L_1\geq L_2$, and 
by $M_1$, $M_2$ and $M_3$ the 
respective minimal labels on the three chains linking them, 
with $M_1\geq M_2 \geq M_3$, the length we are now interested in is nothing but 
the second minimum $M_2$ so that, integrating over the other variables,
we are now led to compute
\eqn\gfncc{\eqalign{& \! {2\over \epsilon^2} 
\int_0^{M_2} \!dM_3\int_{M_2}^\infty \!dM_1
\int_{M_1}^\infty \!dL_2 \int_{L_2}^\infty \!dL_1  \ 
3\, {\tilde {\cal C}}(L_1,M_1) {\tilde {\cal C}}(L_2,M_1)
\times 3\, {\tilde {\cal C}}(L_1,M_2) {\tilde {\cal C}}(L_2,M_2)\cr
& \hskip 8.cm \times 3\, {\tilde {\cal C}}(L_1,M_3) {\tilde {\cal C}}(L_2,M_3)\Big\vert_{M_2=L}
\cr
& ={2 \over \epsilon^2}{3\over {\cal F}'(L)^2}
\int_{0}^{L} \!dM_3 {3\over {\cal F}'(M_3)^2} 
\int_{L}^\infty \!dM_1 {3\over {\cal F}'(M_1)^2} 
\int_{M_1}^\infty \!dL_2 {\cal F}'(L_2)^3  
\int_{L_2}^\infty \!dL_1 {\cal F}'(L_1)^3
\cr
& = {{\cal J}(L)\over \epsilon^2} \quad \quad {\rm with}\quad  
{\cal J}(L)=-{1\over 2} {\cal H}(L)\, {\cal F}'(2L) \cr}
}
with ${\cal H}$ as in \compactrho. 
\fig{Plots of the cumulative probability distribution $\sigma_2(r)$ and the associated
probability density $\sigma_2'(r)$ for the rescaled half-length $r$ of
the ``second shortest" non-contractible loop passing via the origin of
a large pointed toroidal quadrangulation.}{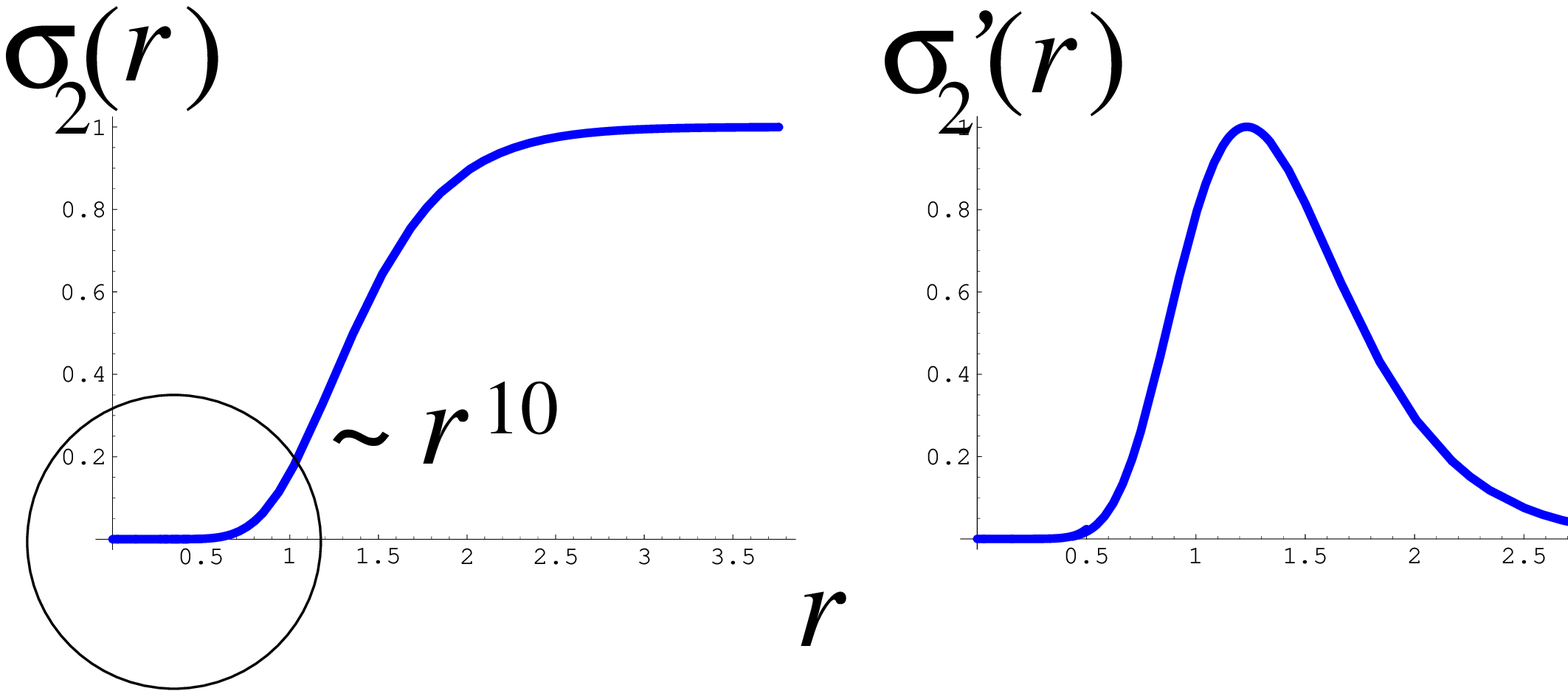}{12.cm}
\figlabel\sigmatwoplots
Repeating the above arguments, we get the {\it probability}
\eqn\sigmatwolaw{\eqalign{\sigma_2(r)& ={96\over \pi}\int_0^{\infty}
d\xi\, {1\over -{\rm i}\xi} \exp(-\xi^2) 
\left({\cal J}(\sqrt{-{\rm i}\xi}r)-{\cal J}(\sqrt{{\rm i}\xi}r)\right)\cr}}
{\it that, in the set of all non-contractible loops passing via the origin, 
the ``second shortest" loop, i.e.\ the shortest among loops not homotopic 
to any power of the true shortest one, has a length smaller than 
$2 r n^{1/4}$ in the ensemble of pointed bipartite quadrangulations of genus 
$1$ with fixed size $n$}, in the limit $n\to\infty$. The cumulative probability 
distribution $\sigma_2(r)$ and the associated probability density 
$\sigma_2'(r)$ are plotted in fig.~\sigmatwoplots.
For small $r$, we have the expansion
\eqn\expansigma{\sigma_2(r)={11043\ r^{10}\over 5096 \sqrt{\pi}}
+{\cal O}(r^{14})\ .}

\fig{A schematic picture of a toroidal quadrangulation with (a)
a small non-contractible loop or (b) two non-homotopic
small non-contractible loops. The first case occurs with a probability
decaying as $n^{-1/2}$ and the second with a probability
decaying as $n^{-3/2}$.}{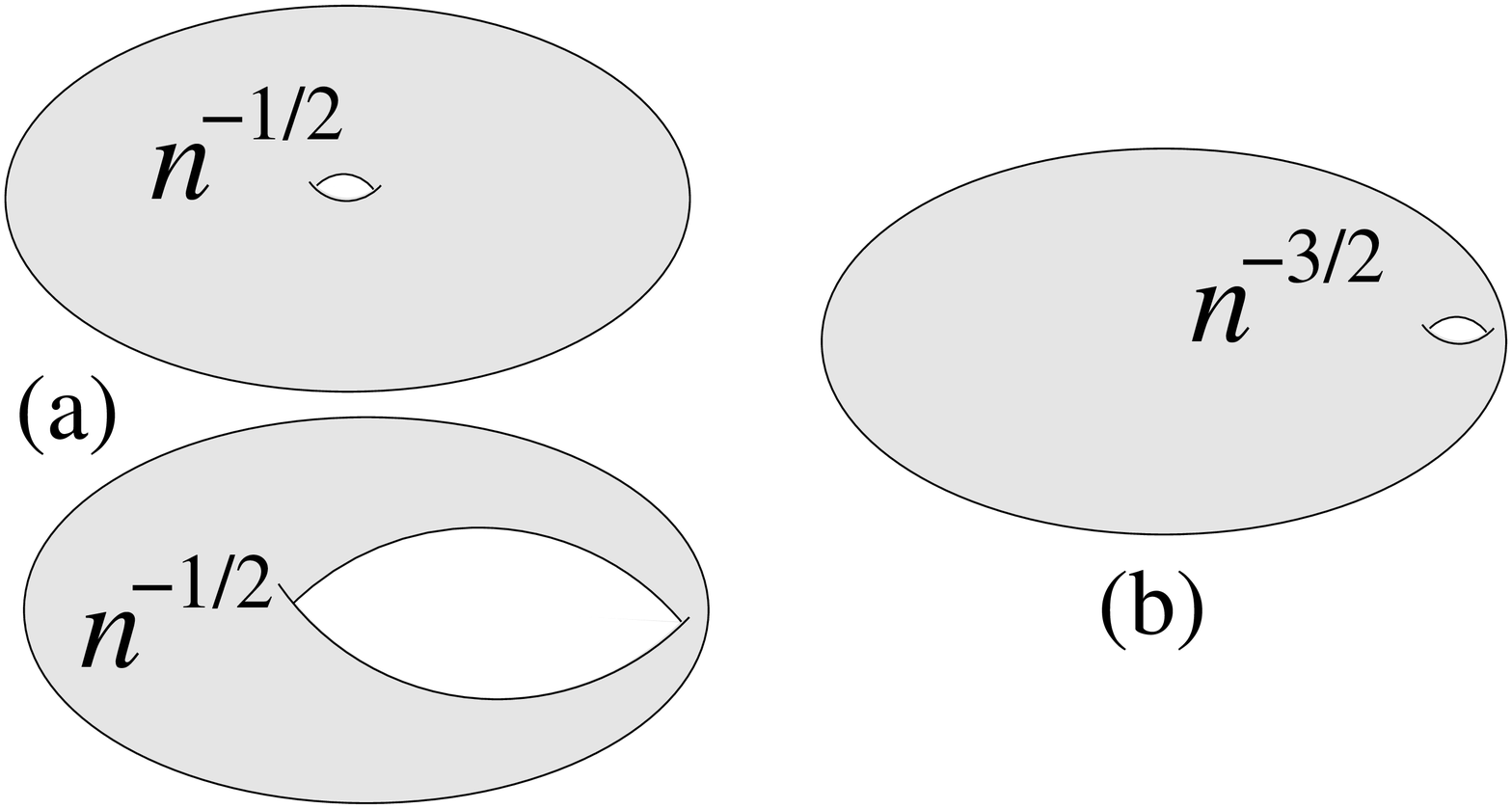}{10.cm}
\figlabel\smallcycles
It is interesting to comment on the small $r$ behavior of 
both $\sigma(r)$ and $\sigma_2(r)$ which informs us on
the regime of large but finite values of $\ell$, namely: 
$1\ll \ell \ll n^{1/4}$. 
For such a finite value of $\ell$, the {\it average 
number of vertices} in a random quadrangulation giving rise to a shortest
non-contractible loop of length smaller than $2\ell$ scales as
\eqn\scalnumb{n \sigma(r) \sim n \times r^6 = 
n \left({\ell\over n^{1/4}}\right)^6= {\ell^6\over \sqrt{n}}}
where we dropped the multiplicative prefactor. In particular, this
quantity tends to $0$ as $n^{-1/2}$ for large $n$. This is to be
contrasted with, for instance, the average number 
($\propto n (\ell/n^{1/4})^4=\ell^4$) 
of vertices at a finite distance less than $\ell$ from the origin of a 
pointed planar quadrangulation \GEOD\ which remains finite at large $n$.
The explanation for this phenomenon is that a prerequisite for 
the shortest non-contractible loop passing via the origin to be 
finite is that the smallest non-contractible loop in the map be itself
finite, while \scalnumb\ essentially counts the vertices lying
at a finite distance from this non-contractible smallest loop. The
$n$ dependence in \scalnumb\ is compatible with a probability of having a 
{\it finite} smallest non-contractible loop decaying as $n^{-1/2}$ 
in the set of bipartite quadrangulations of genus $1$ and fixed large size $n$
(see fig.~\smallcycles-(a) for an illustration). 
Similarly, from the small $r$ behavior of $\sigma_2(r)$, the number
of vertices giving rise to a second-shortest loop of finite length
smaller than $2\ell$ behaves as
\eqn\scalnumbbis{n \sigma_2(r) \sim n \times r^{10} = 
n \left({\ell\over n^{1/4}}\right)^{10}= {\ell^{10}\over n^{3/2}}\ .}
We now interpret the $n$ dependence 
in \scalnumbbis\ as measuring, in the set of 
bipartite quadrangulations with genus $1$ and fixed size $n$, 
the probability that a quadrangulation has its two 
non-homotopic smallest non-contractible loops finite: this probability 
tends to $0$ as $n^{-3/2}$ (see fig.~\smallcycles-(b) for an illustration). 

These scaling behaviors are not surprising since, in case (a) of
fig.~\smallcycles, the quadrangulation may essentially be viewed as 
a planar quadrangulation with two marked points which are glued together
to create a handle, while in case (b), the quadrangulation is essentially 
a planar quadrangulation with one marked point. 
Recall that the number of planar quadrangulations of size $n$ with a 
marked vertex is \TUT
\eqn\qzero{{3^n\over 2n}{{2n\choose n}\over n+1} \sim {12^n\over \sqrt{\pi} 
n^{5/2}}}
while that with two marked vertices is (asymptotically) $n$ times bigger. 
Dividing by the number of total number 
$Q^{(1)}_\to\vert_{g^n}/(4n)\sim 12^n/(96 n)$ of quadrangulations
of genus $1$ gives a ratio scaling precisely as $n^{-1/2}$ for
two marked points and $n^{-3/2}$ for one.

To summarize, we deduce a contrario that a generic bipartite 
quadrangulation of genus $1$
has both its smallest cycles of order $n^{1/4}$ so that its topology
remains that of a genus $1$ surface in the scaling limit.

\newsec{Two-point function for large toroidal maps}
\fig{The ``diagram" enumerating generic configurations contributing to the 
dominant singularity of $Z_\ell$.
Each edge of the diagram must be replaced by a propagator $K$ with 
the indices of its endpoints. We must then sum over
$\ell_1$, $\ell_2$ and $\ell_3$, keeping $\ell$ 
fixed.}{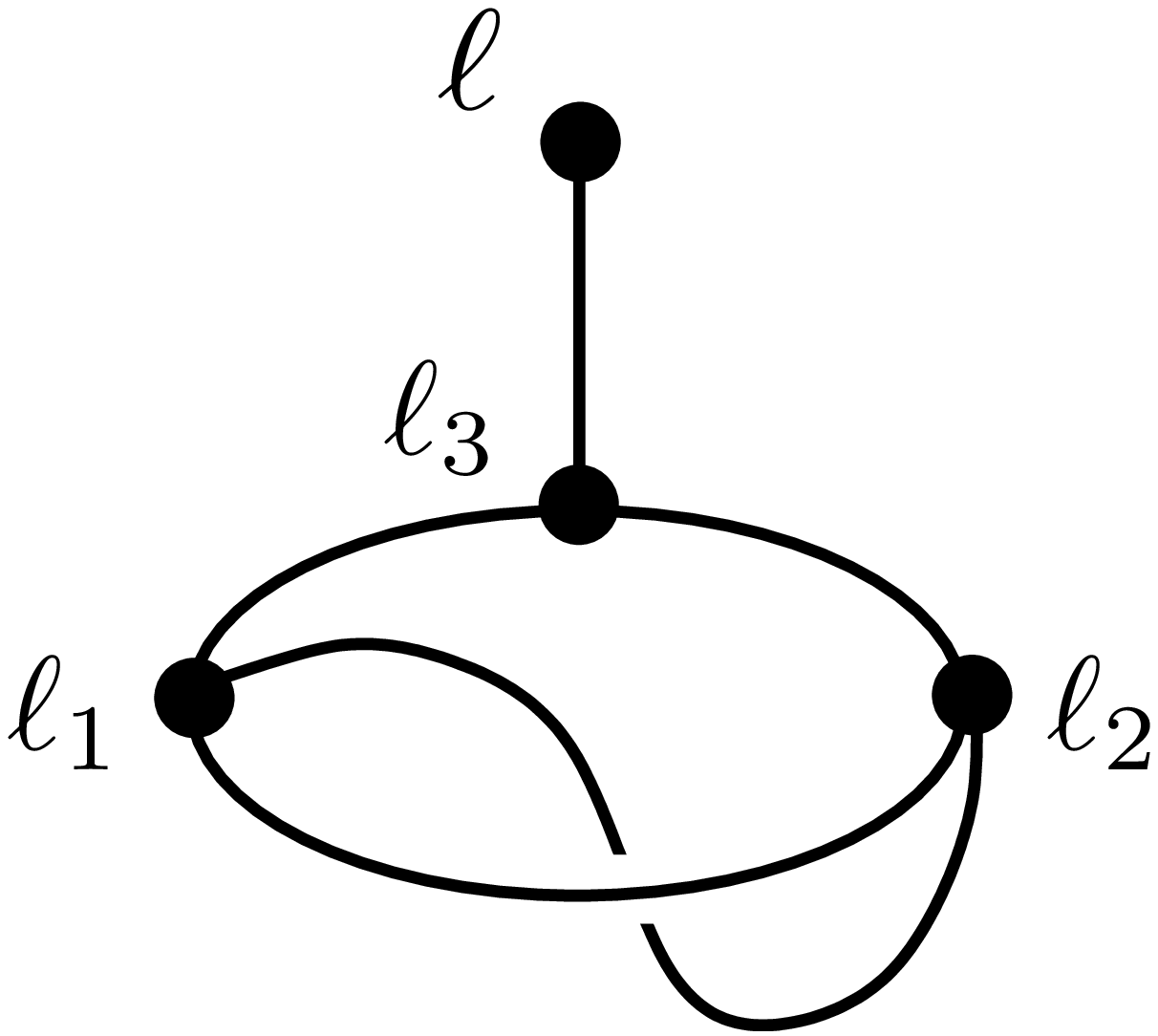}{6.cm}
\figlabel\diagram
The most natural measure of the distance statistics is via the two-point
function which measures the density of points at a fixed distance
from a given origin in the map. More precisely, we may decide to
enumerate pointed bipartite quadrangulations of genus $1$ with now {\it a 
marked vertex at distance $\ell$ ($\ell\geq 1$) from the origin}.
From the Marcus-Schaeffer bijection, this amounts to enumerating
well-labeled $1$-trees with a marked vertex with label $\ell$. 
As before, it is simpler to consider the generating function $Z_\ell$ for 
{\it almost} well-labeled $1$-trees with a marked vertex with label $\ell$.  
From the Marcus-Schaeffer bijection, $Z_\ell$ is now the generating function 
for pointed bipartite quadrangulations of genus $1$ with a marked vertex at 
distance {\it less than or equal to} $\ell$.

The dominant singularity of $Z_\ell$ comes from quadrangulations
leading to well-labeled $1$-trees with a generic backbone and, moreover, 
with a marked 
vertex with label $\ell$ lying outside their skeleton. The shortest path 
in the $1$-tree from the marked vertex to the skeleton defines a chain 
whose endpoint on the skeleton has label, say $\ell_3$, and is
generically different from the two vertices of degree $3$ in the skeleton,
with labels $\ell_1$ and $\ell_2$. The dominant singularity of $Z_\ell$ 
is therefore that of the ``diagram" of fig.~\diagram, whose value is 
obtained by replacing each edge by the corresponding propagator $K$ 
and by summing over $\ell_1$, $\ell_2$ and $\ell_3$, keeping $\ell$ fixed, 
namely
\eqn\equivZell{\sum_{\ell_1\geq 1}\sum_{\ell_2 \geq 1}\sum_{\ell_3\geq 1}
K_{\ell,\ell_3} K_{\ell_1,\ell_3} K_{\ell_2,\ell_3} \left(K_{\ell_1,\ell_2}
\right)^2\ .}
Setting $\ell= L/\sqrt{\epsilon}$ and turning to rescaled variables in
the sums, we get a dominant singularity
\eqn\domsing{\eqalign{Z_\ell &\sim {1\over \epsilon^4} 
\int_0^\infty dL_1 \int_0^\infty dL_2 \int_0^\infty dL_3
\ \rho(L,L_3)\rho(L_1,L_3)\rho(L_2,L_3)\left(\rho(L_1,L_2)\right)^2\cr
& = {{\cal F}^{(1)}(L)\over \epsilon^4}\cr}}
with a new scaling function ${\cal F}^{(1)}(L)$ which fully characterizes
the two-point distance statistics in large toroidal maps. The above integrals
can be performed explicitly and involve $12$ sectors for the determination 
of the minimum in the formula \compactrho\ for $\rho$ ($2$ relative
positions for $L$ and $L_3$ $\times$ $6$ relative positions
for $L_1$, $L_2$ and $L_3$), reducible by symmetry $L_1\leftrightarrow L_2$
to $6$ sectors. After some cumbersome but straightforward 
calculations, we find the relatively simple expression
\eqn\resu{\eqalign{{\cal F}^{(1)}(L)& = 
{\cal A}_0(L)+L\, {\cal A}_1(L)+L^2 {\cal A}_2(L)\cr
{\rm with}&\quad {\cal A}_0(L)={1\over 768}{238+151 \cosh(\sqrt{6}L)+
\cosh(2\sqrt{6}L)\over \sinh^4\left(\sqrt{{3\over 2}} L\right)}
\cr &\quad {\cal A}_1(L)=-{5\over 2048 \sqrt{6}}{100\sinh(\sqrt{6}L)+31
\sinh(2\sqrt{6}L)\over \sinh^6\left(\sqrt{{3\over 2}}L\right)}
\cr &\quad {\cal A}_2(L)=-{75\over 1024}{3+2\cosh(\sqrt{6}L)\over
\sinh^6\left(\sqrt{{3\over 2}}L\right)} 
\ .\cr}}
This explicit formula constitutes the main result of this paper, with
${\cal F}^{(1)}$ playing for the toroidal topology the same role as
${\cal F}$ for the spherical case. 
It is worth noting that, from their explicit expressions, the 
three functions ${\cal A}_0$, ${\cal A}_1$ and ${\cal A}_2$,
satisfy the remarkable relation 
\eqn\propert{{\cal A}_0''-4 {\cal A}_1'+20{\cal A}_2=0}
so that we may write in all generality the parametrization
\eqn\param{{\cal A}_0=4 \alpha_0\ , \quad {\cal A}_1=\alpha_0'+5 \alpha_1\ ,
\quad {\cal A}_2=\alpha_1'\ .}
In other words, the scaling function ${\cal F}^{(1)}$ may be written as
\eqn\parambis{\eqalign{{\cal F}^{(1)}(L)& =4 {\cal M}(L)+L {\cal M}'(L) \cr
{\rm with}&\ {\cal M}(L)=\alpha_0(L)+L \alpha_1(L)\cr}}
in terms of a simpler function ${\cal M}$. We have no explanation
for this particular form nor for the meaning of ${\cal M}(L)$ itself. 
From \resu, we have explicitly
\eqn\valalph{\eqalign{\alpha_0(L)& = {1\over 3072}{238+151\cosh(\sqrt{6}L)+
\cosh(2\sqrt{6}L)\over \sinh^4\left(\sqrt{{3\over 2}} L\right)}
\cr \alpha_1(L) &= {25\over 512} \sqrt{{3\over 2}}
{\cosh\left(\sqrt{{3\over 2}}L \right)\over 
\sinh^5\left(\sqrt{{3\over 2}}L\right)}\ .\cr}}
At small $L$, although ${\cal A}_0$, ${\cal A}_1$ and ${\cal A}_2$ diverge
respectively as $1/L^4$, $1/L^5$ and $1/L^6$, a number of cancellations 
ensure that  ${\cal F}^{(1)}(L)$ vanishes as
\eqn\smallLFone{{\cal F}^{(1)}(L)\sim {L^4\over 896}\ .}
We deduce from this formula that for a large, but {\it finite} $\ell$, i.e.\ 
in the so-called {\it local limit}, we have
\eqn\loclimZ{Z_\ell \sim {L^4\over 896 \epsilon^4} = 
{\ell^4 \over 896 \epsilon^2}
\quad {\rm and\ therefore}
\quad Z_\ell\vert_{g^n} \sim {12^n \over 896}\ \ell^4 }
at large $n$. Normalizing by the number $Q^{(1)}_\bullet\vert_{g^n}
\sim 12^n/96$ 
of pointed quadrangulations, we find that the {\it average number} 
$\langle V_\ell\rangle_\bullet^{(1)}$ {\it of vertices} at a distance 
less than or equal to $\ell$ of the origin in
pointed bipartite quadrangulations of genus $1$ behaves, for large but 
finite $\ell$, as
\eqn\avever{\langle V_\ell\rangle_\bullet^{(1)}\sim {3\over 28} \ell^4\ .} 
This is exactly the result found for planar quadrangulations, i.e.\
in the spherical case $h=0$. As might be expected, the local limit is 
totally insensitive to the genus. The presence of handles may generically 
be felt only by traveling along distances of order $n^{1/4}$ in the map.
\fig{Plots of the cumulative probability distribution 
$\Phi^{(1)}(r)$ and the associated
probability density $\Phi^{(1)}{}'(r)$ for the rescaled distance $r$ 
from a uniformly chosen random vertex to the origin of a 
large pointed toroidal quadrangulation.}{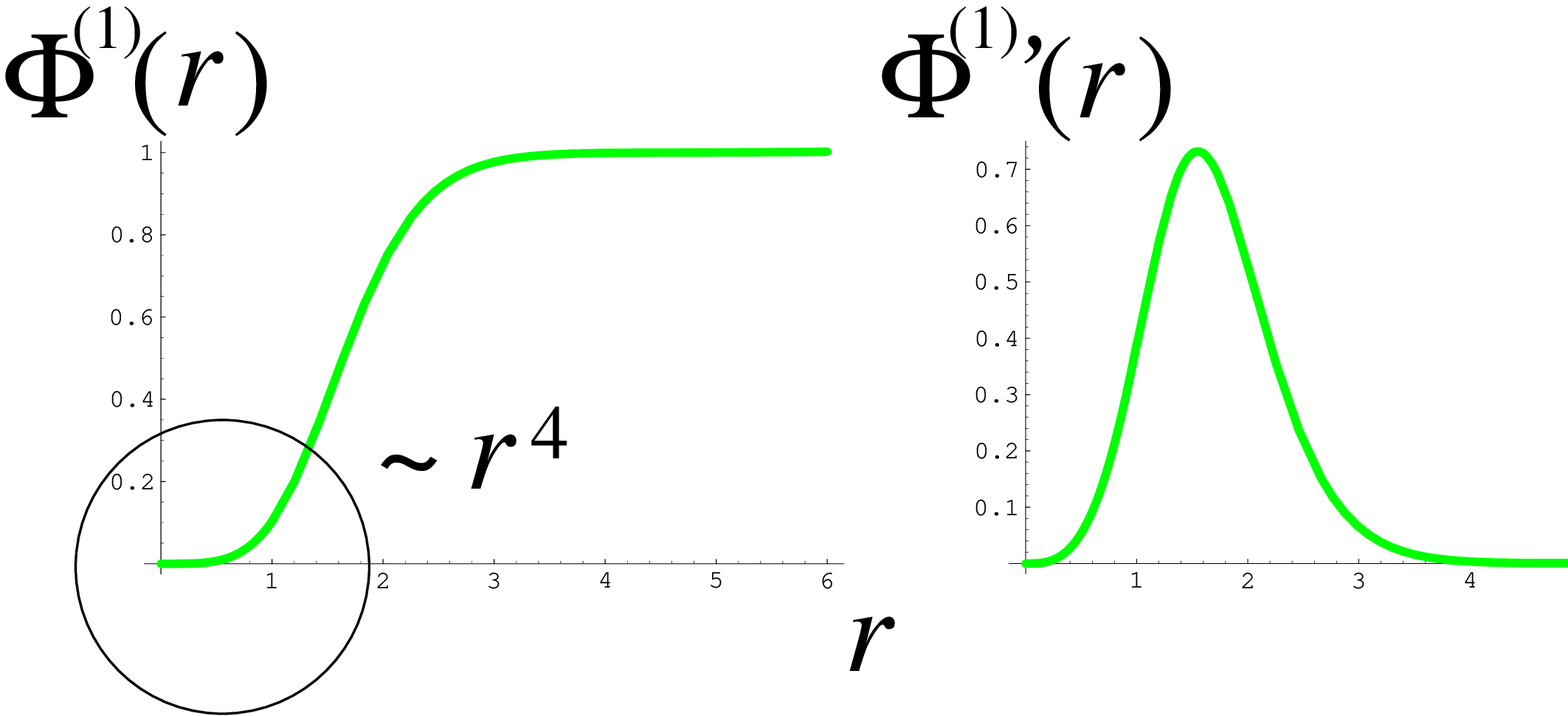}{12.cm}
\figlabel\phioneplots
\fig{Comparison of the probability densities $\Phi^{(1)}{}'(r)$ and 
$\Phi^{(0)}{}'(r)$ for the rescaled distance $r$ from a uniformly chosen 
random vertex to the origin in large pointed toroidal and planar 
quadrangulations respectively. We plotted on the right the difference
$\Phi^{(1)}{}'-\Phi^{(0)}{}'$.}{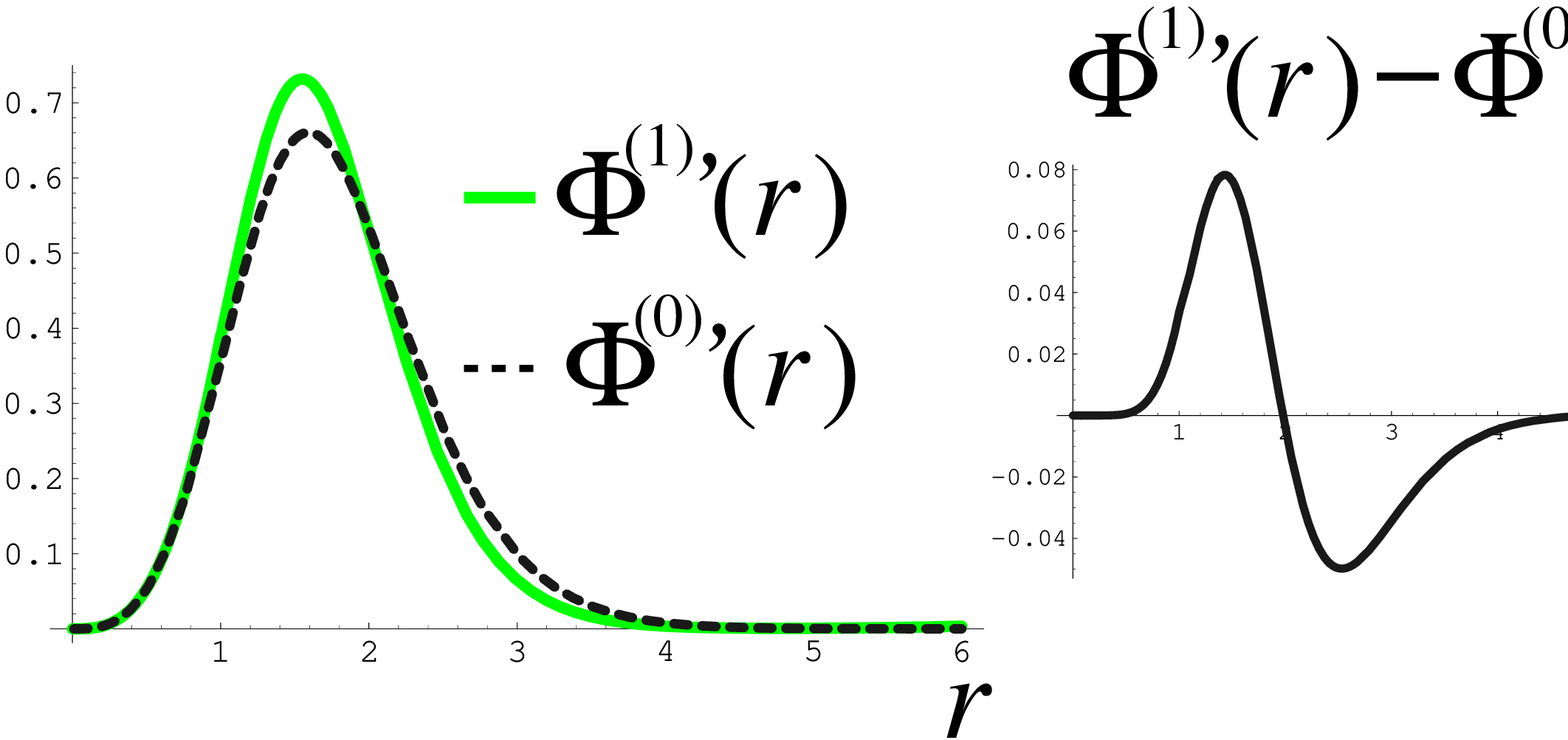}{12.5cm}
\figlabel\compplots
Alternatively, we may consider the scaling limit of pointed bipartite 
quadrangulations of genus $1$ with fixed, large size $n$, but now with 
$\ell\propto n^{1/4}$. The probability 
$\Phi^{(1)}(r)$ that a vertex picked uniformly at random in the quadrangulation
be at a distance less than $\ell=r n^{1/4}$ from the origin
is given by 
\eqn\phione{\eqalign{\Phi^{(1)}(r)
& = {96\over \pi}
\int_{-\infty}^\infty d\xi\, {1\over {\rm i} \xi^3} \exp(-\xi^2) 
{\cal F}^{(1)}(\sqrt{-{\rm i}\xi}\, r)\cr 
& = {3\, r^4 \over 28}+{96\over \pi}
\int_{-\infty}^\infty d\xi\, {1\over {\rm i} \xi^3} \exp(-\xi^2) 
\left({\cal F}^{(1)}(\sqrt{-{\rm i}\xi}\, r) - {(\sqrt{-{\rm i}\xi}\, r)^4
\over 896}\right) \cr 
&= {3\, r^4 \over 28}+{96\over \pi}
\int_0^\infty d\xi\, {1\over {\rm i} \xi^3} 
\exp(-\xi^2) \left({\cal F}^{(1)}(\sqrt{
-{\rm i}\xi}\, r) - {\cal F}^{(1)}(\sqrt{{\rm i}\xi}\, r)\right)\cr}}
where we singled out the contribution from the first term in the small $r$ 
expansion of ${\cal F}^{(1)}(r)$, corresponding precisely to the 
local limit, as it is proportional to the improper integral 
$\int_{-\infty}^\infty (d\xi/\xi) \exp(-\xi^2)$. Its value $-{\rm i}\pi$ 
of this integral is 
dictated by the change of variable from $g$ to $\xi$. The integrals in the 
second and third lines of \phione\ are now convergent integrals. Expanding 
further at small $r$ gives
\eqn\expanphi{\Phi^{(1)}(r)={3 r^4\over 28}-{15 r^{10}\over 1456 \sqrt{\pi }}
+{1242135 r^{14}\over 506970464 \sqrt{\pi }}+{\cal O}\left(r^{18}\right)}
with, as discussed above, the same leading term $\propto r^4$ as for
planar quadrangulations, but now a first negative correction of order $r^{10}$
instead of $r^{8}$ for the planar two-point function. The two-point function
therefore increases faster at small $r$ for toroidal maps than for
spherical ones.
The probability distribution $\Phi^{(1)}(r)$ and the associated 
probability density $\Phi^{(1)}{}'(r)$ are plotted in fig.~\phioneplots.
A comparison with the corresponding genus $0$ probability density
$\Phi^{(0)}{}'(r)$, as computed in ref.~\GEOD, is displayed in fig.~\compplots.

\newsec{Conclusion}

In this paper, we have derived explicit expressions for a number of 
probability distributions characterizing the distance statistics 
of large toroidal maps. These distributions, obtained here in the context of
bipartite quadrangulations, are expected to be universal 
(up to a non-universal global rescaling of $r$) and describe 
the distance statistics in more general ensembles of large toroidal maps 
in the universality class of pure gravity, such has maps with prescribed
face degrees, possibly equipped with non-critical statistical models. 

Our main result is the explicit form \resu\ for the two-point scaling
function ${\cal F}^{(1)}$, which is the genus $1$ counterpart of  
spherical two-point scaling function ${\cal F}$ of eq.~\expanRell. 
In the same way as ${\cal F}$ satisfies the non-linear differential
equation \eqforF, we may wonder if ${\cal F}^{(1)}$ itself obeys some 
simple differential equation, possibly involving ${\cal F}$ as a source. 
Our method, which consisted in a direct computation of ${\cal F}^{(1)}$, 
did not allow us to find such an equation. 

Among possible extensions of our work, let us mention the computation of the
toroidal three-point function or more simply, that of a more involved
two-point function now measuring the "second-shortest length 
between two points", i.e.\ the length of any shortest path among those paths
not homotopic to the true geodesic. There seems to be no fundamental 
obstacle to the derivation of these laws but the calculations may rapidly 
become heavy.

Another natural extension if of course that of maps with genus $h>1$.
Here a more fundamental obstacle occurs since, when $h$ becomes large, 
we have to deal with a large number of diagrams, each involving
a large number of propagators, which
makes in practice our method unadapted. Alternatively,
one may hope for the existence of a hierarchy of equations
satisfied by the higher genus two-point functions, whose
discovery would be a promising step in the quest for these 
universal scaling functions. 

\listrefs
\end